\documentclass[twocolumn,english]{revtex4-1}
\usepackage[T1]{fontenc}
\usepackage[latin9]{inputenc}
\usepackage{color}
\usepackage{amsmath}
\usepackage{graphicx}
\usepackage{esint}

\makeatletter

\usepackage{babel}

\usepackage{babel}

\usepackage{babel}

\makeatother

\usepackage{babel}
\begin{document}
\title{Hydrodynamic transport and violation of the viscosity-to-entropy ratio
bound in nodal-line semimetals}
\author{Sang Wook Kim, Geo Jose, and Bruno Uchoa}
\affiliation{Center for Quantum Research and Technology, Department of Physics
and Astronomy, University of Oklahoma, Norman, Oklahoma 73019, USA }
\date{\today}
\begin{abstract}
The ratio between the shear viscosity and the entropy $\eta/s$ is
considered a universal measure of the strength of interactions in
quantum systems. This quantity was conjectured to have a universal
lower bound $(1/4\pi)\hbar/k_{B}$, which indicates a very strongly
correlated quantum fluid. By solving the quantum kinetic theory for
a nodal-line semimetal in the hydrodynamic regime, we show that $\eta/s\propto T$
violates the universal lower bound, scaling towards zero with decreasing
temperature $T$ in the perturbative limit. We find that the hydrodynamic
scattering time between collisions is nearly temperature independent,
up to logarithmic scaling corrections, and can be extremely short
for large nodal lines, near the Mott-Ragel-Ioffe limit. Our finding
suggests that nodal-line semimetals can be very strongly correlated
quantum systems.
\end{abstract}
\maketitle

\section{Introduction}

Hydrodynamics describes the behavior of quantum fluids in the regime
where the relaxation of electrons is dominated by collision among
the quasiparticles. This theory describes long wavelength deviations
from local thermal equilibrium, when transport is dominated by conservation
laws \cite{Hartnoll}. Since the time between collisions is the shortest
time scale in the problem, the electrons exchange momentum faster
than they can relax to phonons or disorder. That leads to universal
behavior in the form of a slow diffusion of densities and to viscous
flow. This framework has been successfully applied to a variety of
different systems, ranging from strong coupling gauge theories with
holographic duals \cite{Maldacena}, quark-gluon plasma \cite{Shuryak},
cold-atoms systems \cite{Joseph,Clancy}, thin wires \cite{Moll},
and graphene \cite{Muller,Baldurin,Crossno}.

The shear viscosity measures the longitudinal resistivity to transverse
gradients in the velocity of a fluid. It has been conjectured by Kovtun
\emph{et al.} \cite{Kovtun} that quantum systems have a universal
lower bound for the ratio between the sheer viscosity and the entropy,
\begin{equation}
\frac{\eta}{s}\geq(1/4\pi)\hbar/k_{B}.\label{1}
\end{equation}
The equality was found in an infinitely strongly coupled field theory
and has been associated with ``perfect fluids,'' systems that are
so strongly interacting that they can display quantum turbulence \cite{Muller,Shavit}.
This ratio is widely believed to be a proxy for the strength of interactions
in many classes of quantum systems, including relativistic, non-relativistic
systems and Plankian metals \cite{Patel}, which entirely lack quasiparticles.

By dimensional analysis, the shear viscosity $\eta\sim F\tau$, where
$F$ is the free energy and $\tau$ is the relaxation time \cite{Zaanen2,Zaanen}.
In hydrodynamic relativistic systems, the free energy is mostly entropic,
$F\sim sT$. In the absence of screening, the scattering time due
to Coulomb interactions is $\tau\sim\hbar/(k_{B}T)$, and hence $\eta/s\sim\hbar/k_{B}$,
with a prefactor of order unity. In general, screened electronic quasiparticles
are long lived and typically lead to high viscosity in quantum fluids.
In Fermi liquids, the free energy is dominated by the Fermi energy
$E_{F}$ at low temperature, whereas $\tau\propto T^{-2}$. The ratio
$\eta/s\sim\hbar/k_{B}(E_{F}/T)^{3}$ for $T<E_{F}$ \cite{Abrikosov},
saturating to a constant $\eta/s\sim\hbar/k_{B}$ at $T>E_{F}$, above
the conjectured universal lower bound.

\begin{figure}[b]
\begin{centering}
\includegraphics[scale=0.3]{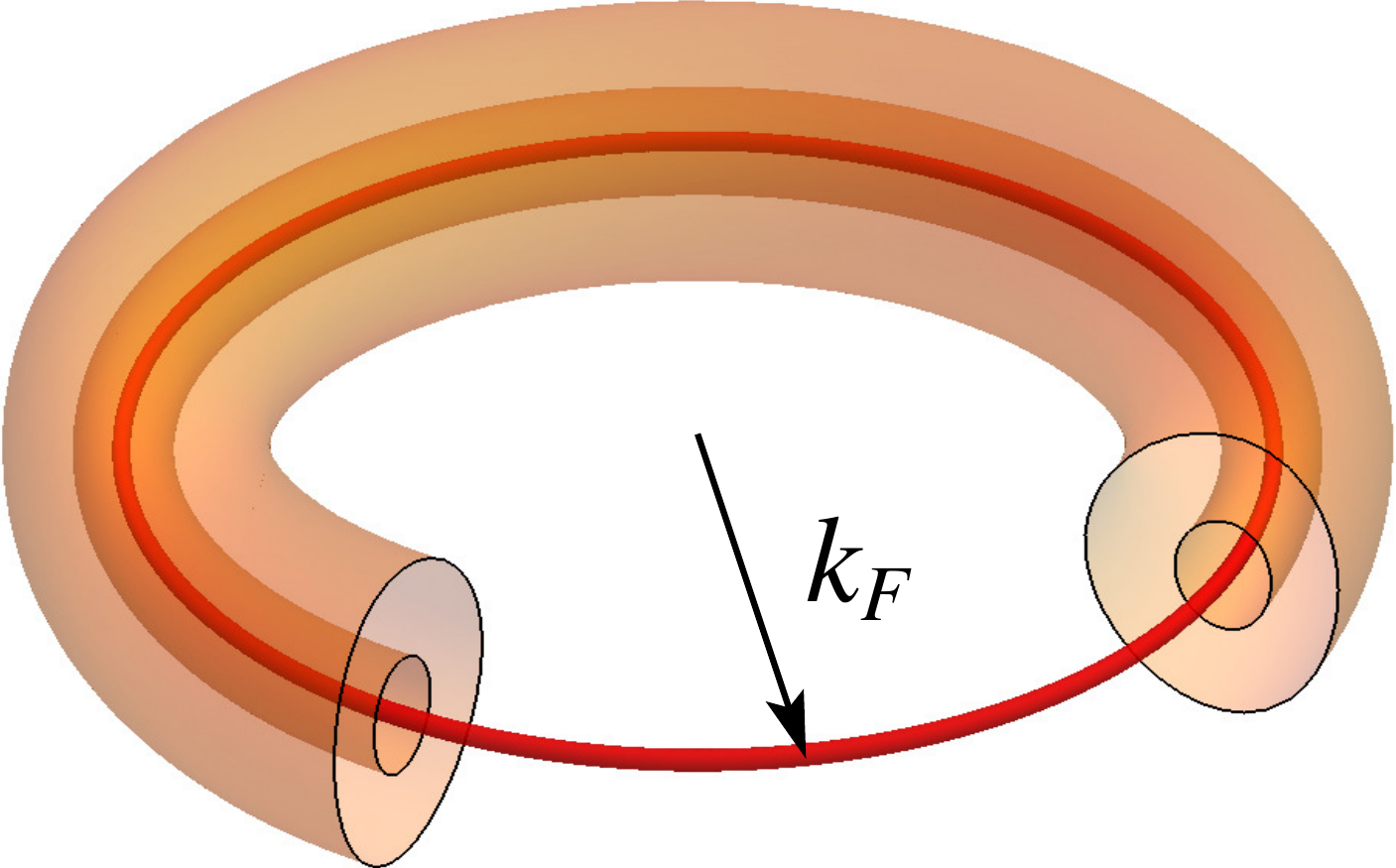} 
\par\end{centering}
\caption{{\small{}\label{fig:Fermi-surface}Fermi surface of a NLSM. Massless
quasiparticles disperse linearly away from a nodal line (red) with
radius $k_{F}$. The toruses enclosing the nodal line are finite energy
surfaces. The outer shell with energy $k_{B}\Lambda_{T}$ sets the
ultraviolet temperature cut-off of the theory.}}
\end{figure}

Violations of the universal bound were found before in some strongly
interacting conformal field theories \cite{counter1,counter2,counter3}
and holographic gravity models \cite{Hartnoll2,Alberte}, and were
predicted near a superfluid transition \cite{Gochan}. In quantum
materials, it has been recently suggested that anisotropic Dirac fermions
found at a topological Lifshitz transition, where two Dirac cones
merge \cite{Adroguer}, violate the proposed lower bound in the non-perturbative
regime of interactions \cite{Link}. Coulomb interactions, nevertheless,
were more recently shown to restore the isotropy of the Dirac cone
near the fixed point of that problem \cite{Kotov2}, effectively reinstating
a lower bound. The extent to which the universal lower bound is violated
(or not) in that problem requires a closer examination.

In this paper, we show that in nodal systems where the density of
states vanishes along a Fermi line, such as in nodal-line semimetals
(NLSMs) \cite{Burkov,Mullen,Yang,Rappe,Weng,Yu,Heykikila,Chen,Xie,Bian1,Bian2,Song,Fu},
the ratio between the shear viscosity and the entropy strongly violates
the conjectured lower bound, scaling towards zero with decreasing
temperature in the perturbative regime, 
\begin{equation}
\frac{\eta}{s}\propto\frac{\hbar}{k_{B}}\frac{k_{B}T}{\alpha^{2}v_{F}k_{F}}\sim T\tau,\label{eta/s}
\end{equation}
where $k_{F}$ is the radius of the nodal line, $v_{F}$ is the Fermi
velocity of the quasiparticles, and $\alpha=e^{2}/v_{F}$ is the fine
structure constant. This is the main result of the paper.

In the absence of screening, the scattering rate $\tau^{-1}$ is set
by the volume of the phase space available for collisions. Due to
the lack of dispersion along the line, as illustrated in Fig. \ref{fig:Fermi-surface},
there is no energy cost for the quasiparticles to scatter in that
direction, even at zero temperature. From this phase space argument,
the scattering time is hence temperature independent, scales inversely
with the length of the nodal line and can be extremely short for large
nodal lines, possibly close to the Mott-Ragel-Ioffe limit \cite{Ioffe}.
We find that 
\begin{equation}
\tau\sim\frac{\hbar}{\alpha^{2}v_{F}k_{F}},\label{eq:tau}
\end{equation}
with additional logarithmic scaling corrections in temperature in
the perturbative regime.

We confirm that result by calculating the longitudinal conductivity
in the collision dominated regime ($\omega\ll\tau^{-1}$), 
\begin{equation}
\sigma(\omega,T)\propto\frac{e^{2}}{h}\frac{k_{B}T}{\alpha^{2}v_{F}}\sim\frac{e^{2}}{h}k_{F}(k_{B}T)\frac{\tau}{\hbar},\label{eq:sigma0}
\end{equation}
which is indicative of insulating behavior. We note that in Weyl semimetals,
the dc conductivity $\sigma\propto T^{2}\tau$ also scales linearly
with temperature (since $\tau\propto1/T$, as in graphene \cite{note5,Parameswaran}),
although reflecting a completely distinct behavior in the scaling
of the scattering rate, and hence in the viscosity-to-entropy ratio.
We conclude that the violation of the bound due to an unusually short
and nearly temperature-independent scattering time suggests that NLSMs
can be extremely correlated quantum systems.

In the following, we outline the structure of the paper. In Sec. II,
we derive the quantum kinetic equation. In Sec. III, we calculate
the conductivity in the hydrodynamic regime, including a discussion
on many-body effects through a renormalization group analysis. In
Sec. IV, we calculate the shear viscosity and demonstrate the violation
of the viscosity-to entropy ratio bound. Finally, in Sec. V, we discuss
experimental implications of this result.

\section{Quantum kinetic equation}

We adopt the low-energy Hamiltonian of a NLSM that is described by
a circular nodal line in the $k_{z}=0$ plane. The low-energy quasiparticles
are Dirac fermions located in the vicinity of the nodal line, 
\begin{equation}
\mathcal{H}_{0}(\mathbf{k})=\frac{k_{x}^{2}+k_{y}^{2}-k_{F}^{2}}{2m}\sigma_{x}+v_{z}k_{z}\sigma_{y}\approx v_{F}\delta k_{r}\sigma_{x}+v_{z}k_{z}\sigma_{y},\label{eq:H1}
\end{equation}
where $\delta k_{r}=k_{r}-k_{F}$ is the in-plane momentum away from
the nodal line, and $v_{F}=k_{F}/m$ is the Fermi velocity in the
radial direction and $v_{z}$ along the the $z$ direction. The quasiparticles
interact through the three-dimensional (3D) Coulomb potential 
\begin{equation}
V(q)=4\pi\frac{e^{2}}{q^{2}}\label{Coulomb}
\end{equation}
and disperse linearly near the nodal line.

In the hydrodynamic regime, the particles interact with each other
more quickly than they lose energy to the lattice. The electronic
relaxation is driven by the collision between particles, leading to
local thermalization. The out-of-equilibrium distribution function
of the quasiparticles $f_{\lambda}(\mathbf{k},\mathbf{x},t)$ satisfies
the Boltzmann equation 
\begin{equation}
\left(\frac{\partial}{\partial t}+\mathbf{v}_{\lambda,\mathbf{k}}\cdot\nabla_{\mathbf{x}}+e\mathbf{E}\cdot\nabla_{\mathbf{k}}\right)f_{\lambda}=\mathcal{I}_{\textrm{col}}[f_{\lambda}],\label{eq:BE}
\end{equation}
where $\lambda=\pm1$ for quasiparticles and quasiholes respectively,
and $\mathbf{v}_{\lambda,\mathbf{k}}=\nabla_{\mathbf{k}}\varepsilon_{\lambda,k}$
is the velocity of the quasiparticles, with 
\begin{equation}
\varepsilon_{\lambda,\mathbf{k}}^{0}=\lambda\sqrt{(v_{F}\delta k_{r})^{2}+(v_{z}k_{z})^{2}}\label{epsilon0}
\end{equation}
being the equilibrium energy spectrum. The term $e\mathbf{E}=\partial\mathbf{k}/\partial t$
is the external force driving the system, with $\mathbf{E}$ being
the electric field, and $\mathcal{I}_{\textrm{col}}[f_{\lambda}]$
is the collision integral, which includes all scattering processes
between quasiparticles allowed by Fermi's golden rule. For a nonequilibrium
state, 
\begin{equation}
f_{\lambda}\left(\mathbf{k},\mathbf{x},t\right)=f_{\lambda}^{0}\left(\mathbf{k}\right)+\delta f_{\lambda}\left(\mathbf{k},\mathbf{x},t\right),\label{eq:f}
\end{equation}
where $f_{\lambda}^{0}=[e^{\varepsilon_{\lambda}^{0}\beta}+1]^{-1}$
is the equilibrium Fermi distribution, which solves the Boltzmann
equation in the absence of interactions $(I_{\text{col}}=0$), $\beta=1/k_{B}T$,
and $\delta f_{\lambda}\left(\mathbf{k},\mathbf{x},t\right)$ is the
non-equilibrium correction in linear response to an external perturbation
such as electric field and strain. In general, $\mathcal{I}_{\textrm{col}}\approx\delta f/\tau$,
where $\tau$ is the scattering time between collisions.

\section{Conductivity}

To gain physical intuition in the problem, we derive first the conductivity
and the scattering time for NLSMs. If the system is spatially homogenous,
the non-equilibrium current carried by the quasiparticles in the presence
of an external electric field is
\begin{equation}
\mathbf{J}=e\sum_{\lambda}\int_{\mathbf{k}}\mathbf{v}_{\lambda,\mathbf{k}}f_{\lambda}(\mathbf{k},\omega),\label{eq:J}
\end{equation}
with $\int_{\mathbf{k}}\equiv(2\pi)^{-3}\int\text{d}^{3}k$. In linear
response, where $J_{i}=\sigma_{ij}E_{j}$, the conductivity per spin
is 
\begin{equation}
\sigma_{ij}(\omega,T)=e\sum_{\lambda}\int_{\mathbf{k}}(v_{\lambda,\mathbf{k}})_{i}\frac{\partial}{\partial E_{j}}\delta f_{\lambda}(\mathbf{k},\omega).\label{sigma}
\end{equation}
In leading order and close to equilibrium, the driving force term
on the left-hand side of (\ref{eq:BE}) is
\begin{equation}
-e\mathbf{E}\cdot\nabla_{\mathbf{k}}f_{\lambda}=e\mathbf{E}\cdot\boldsymbol{\phi}_{\lambda}(k)\label{Enabla}
\end{equation}
with $\phi_{\lambda,i}(k)\equiv\beta f_{\lambda}^{0}(1-f_{\lambda}^{0})(v_{\lambda,\mathbf{k}})_{i}.$
The non-equilibrium dispersion can be written in the form
\begin{equation}
\varepsilon_{\lambda,k}=\varepsilon_{\lambda,\mathbf{k}}^{0}+e\mathbf{E}(\omega)\cdot(\nabla_{\mathbf{k}}\varepsilon_{\lambda,k}^{0})g_{\lambda}(k),\label{energy}
\end{equation}
with $g_{\lambda}\left(k\right)$ being some unknown function to be
found from the solution of the kinetic equation, where $k\equiv(\mathbf{k},\omega)$.
With this ansatz, the non-equilibrium correction of the distribution
function assumes the form 
\begin{equation}
\delta f_{\lambda}\left(k\right)=\beta f_{\lambda}^{0}\left(1-f_{\lambda}^{0}\right)e\mathbf{E}\left(\omega\right)\cdot\mathbf{v}_{\lambda,\mathbf{k}}g_{\lambda}(k).\label{deltaf}
\end{equation}
For convenience, we define $\chi_{\lambda,i}\equiv(v_{\lambda,\mathbf{k}})_{i}g_{\lambda}$.
In the collision dominated regime $\omega\ll\tau^{-1}$, the linearized
kinetic equation (2) can be approximately expressed in terms of the
collision operator as
\begin{equation}
\phi_{\lambda,i}=\mathcal{C}\chi_{\lambda,i},\label{phiC}
\end{equation}
where\begin{widetext}
\begin{align}
\mathcal{C}\chi_{\lambda,i} & =\sum_{\lambda_{1}\lambda_{2}\lambda_{3}}\int_{\mathbf{k}_{1}}\int_{\mathbf{k}_{2}}\int_{\mathbf{k}_{3}}\left(2\pi\right)^{4}\delta^{3}\left(\mathbf{k}+\mathbf{k}_{1}-\mathbf{k}_{2}-\mathbf{k}_{3}\right)\delta\left(\varepsilon_{\lambda,\mathbf{k}}^{0}+\varepsilon_{\lambda_{1},\mathbf{k}_{1}}^{0}-\varepsilon_{\lambda_{2},\mathbf{k}_{2}}^{0}-\varepsilon_{\lambda_{3},\mathbf{k}_{3}}^{0}\right)\mathcal{M}_{\lambda\lambda_{1}\lambda_{2}\lambda_{3}}^{\textrm{Col}}f_{\lambda}^{0}f_{\lambda_{1}}^{0}f_{-\lambda_{2}}^{0}f_{-\lambda_{3}}^{0}\nonumber \\
 & \quad\qquad\qquad\qquad\qquad\qquad\times\left[\chi_{\lambda,i}\left(k\right)+\chi_{\lambda_{1},i}\left(k_{1}\right)-\chi_{\lambda_{2},i}\left(k_{2}\right)-\chi_{\lambda_{3},i}\left(k_{3}\right)\right],\label{C}
\end{align}
\end{widetext}with $\mathcal{M}_{\lambda\lambda_{1}\lambda_{2}\lambda_{3}}^{\textrm{Col}}$
being the collision matrix element \cite{Fritz}. For details in the
derivation of the collision term and integration, see Appendix A.

The solution of the Boltzmann equation requires inverting the collision
operator $\mathcal{C}$, which can be done through the standard procedure
\cite{Fritz}. The dominant contribution to the conductivity follows
from the eigenfunctions of the collision operator with the lowest
eigenvalues. In the collinear approximation, where the momenta of
the quasiparticles point in the same direction, the momenta embedded
in the definition of the velocities $\mathbf{v}_{\lambda,\mathbf{k}}$
factor out in the integrand of $\mathcal{C}$, which is proportional
to 
\begin{equation}
\lambda g_{\lambda}\left(k\right)+\lambda_{1}g_{\lambda_{1}}(k_{1})-\lambda_{2}g_{\lambda_{2}}(k_{2})-\lambda_{3}g_{\lambda_{3}}\left(k_{3}\right).\label{eq:Cons}
\end{equation}
The zero modes of the collision operator $\mathcal{C}\chi_{\lambda,i}=0$
in this restricted phase space are
\begin{equation}
g_{1,\lambda}(k)=a^{(e)}(\omega),\label{g1}
\end{equation}
\begin{equation}
g_{2,\lambda}(k)=a^{(\chi)}(\omega)\lambda,\label{g2}
\end{equation}
and
\begin{equation}
g_{3,\lambda}(k)=a^{(p)}(\omega)\varepsilon_{\lambda,k}^{0},\label{g3}
\end{equation}
corresponding to conservation of charge, chirality, and momentum,
respectively.

In the absence of noncollinear processes, those zero modes would produce
infinite conductivity \cite{Fritz}. To account for non-collinear
processes, we express the eigenfunctions of full collision operator
$\mathcal{C}$ that have the lowest eigenvalues in a basis of zero
modes of the collinear regime. We note that due to translational symmetry,
the momentum zero mode is an exact eigenfunction of (\ref{C}), as
can be readily checked \cite{Fritz,Kashuba}. It does not, however,
contribute to the conductivity (\ref{sigma}) due to particle-hole
symmetry at the nodal line. For the same reason, the chiral modes
do not contribute the the charge transport either. We are then left
with the charge zero mode, $\chi_{\lambda,i}(k)=a^{(e)}(\omega)(v_{\lambda,\mathbf{k}})_{i}$,
which provides the only contribution to the conductivity.

We next restore the frequency dependence of the Boltzmann equation,
$\phi_{\lambda,i}=\mathcal{C}\chi_{\lambda,i}+i\omega g_{\lambda}\phi_{\lambda,i}$.
In order to calculate the function $a^{(e)}(\omega)$, we define the
inner product $(a_{\lambda,i},b_{\lambda,i})=\sum_{\lambda,i}\int_{\mathbf{k}}a_{\lambda,i}(\mathbf{k})b_{\lambda,i}(\mathbf{k})$
and set the variational functional 
\begin{equation}
Q\left[a^{(e)}\right]\equiv\left(\chi_{\lambda,i},\phi_{\lambda,i}\right)-\frac{1}{2}\left(\chi_{\lambda,i},\mathcal{C}\chi_{\lambda,i}+i\omega a^{(e)}\phi_{\lambda,i}\right),\label{eq:Q}
\end{equation}
which is to be minimized, $\partial Q/\partial a^{(e)}=0.$ The momentum
integral of the collision operator is performed in the collinear approximation,
where all momenta are nearly parallel to each other. As shown in Appendix
B, this approximation is justified by the fact that for a large nodal
line ($v_{F}k_{F}\gg k_{B}\Lambda_{T}$), the weight of collinear
processes in the collision phase space is logarithmically divergent,
as in the case of two-dimensional (2D) Dirac fermions \cite{Fritz,Muller}.
We also restrict scattering to channels that conserve the number of
particles and holes, which are dominant processes in the collinear
regime.

Combining the solution of Eq. (\ref{eq:Q}) with Eqs. (\ref{sigma})
and (\ref{deltaf}), we obtain the frequency dependent conductivity
in the hydrodynamic regime, 
\begin{equation}
\sigma_{ii}(\omega,T)=\gamma_{i}\frac{e^{2}}{h}Nk_{F}\frac{k_{B}T}{i\omega+\alpha^{2}(T)v_{F}(T)k_{F}c(\gamma)},\label{sigma2}
\end{equation}
where
\begin{equation}
\gamma_{z}=\gamma\equiv v_{z}/v_{F},\qquad\gamma_{i}=\gamma^{-1},\label{gamma}
\end{equation}
for $i=x,y$ and $N$ is the spin degeneracy \cite{note6}. The coefficient
$c(1)\approx1.034$ was numerically extracted from the collision integral
for $N=2$. This value decreases monotonically away from $\gamma=1$.
The functions $\alpha(T)$ and $v_{F}(T)$ are the fine-structure
constant and Fermi velocity, respectively, dressed by interaction
effects.

\subsection{Renormalization group analysis}

As in graphene \cite{Kotov}, Coulomb interactions are marginal and
renormalize the velocity of the quasiparticles in the perturbative
regime. The velocity grows logarithmically with decreasing temperature,
\begin{equation}
v_{F}(T)=v_{F}\left[1+\frac{\alpha}{4}\ln\left(\frac{\Lambda_{T}}{T}\right)\right],\label{eq:v}
\end{equation}
where $k_{B}\Lambda_{T}=v_{F}\Lambda\ll v_{F}k_{F}$ is the ultraviolet
cutoff. The electron charge does not run and the fine-structure constant
is also renormalized,
\begin{equation}
\alpha(T)=\frac{\alpha}{1+\frac{\alpha}{4}\ln\left(\Lambda_{T}/T\right)},\label{alpha}
\end{equation}
and decreases logarithmically at low temperature. The renormalization
group (RG) results mimic the structure of the calculation in graphene.
The details can be found in Appendix C.

\begin{figure}
\begin{centering}
\includegraphics[scale=0.28]{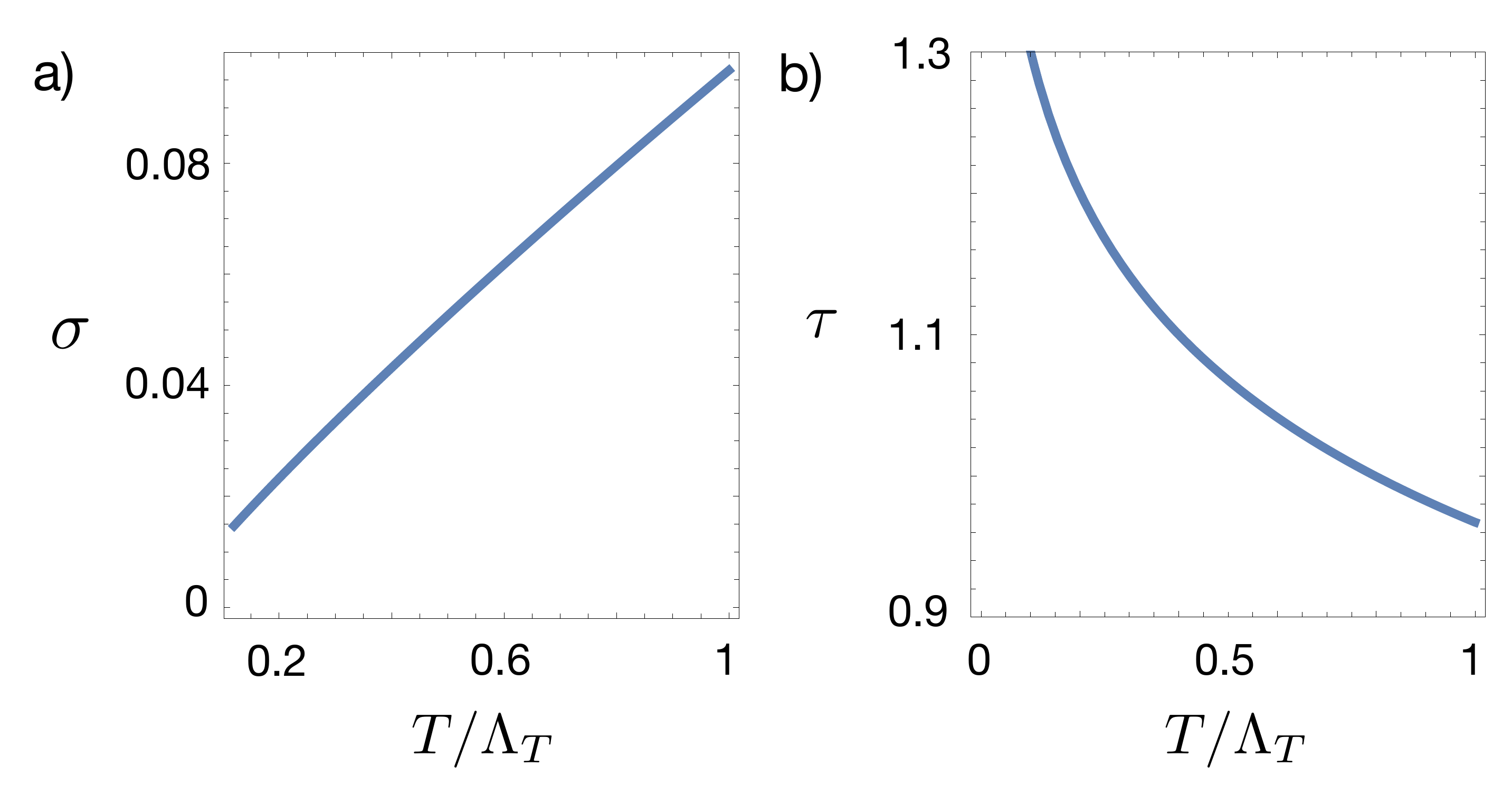}
\par\end{centering}
\caption{{\small{}\label{fig:conductivity}(a) Longitudinal conductivity $\sigma(0,T)$
in units of $e^{2}k_{F}/(h\alpha^{2})$ vs temperature normalized
by the ultraviolet temperature cutoff $\Lambda_{T}$ in the collision
dominated regime, $\omega\ll\tau^{-1}$. The conductivity has a quasi-linear
scaling in the range $T\in[\lambda_{T},T]$, with $\lambda_{T}/\Lambda_{T}\sim0.2$
(see text). (b) Scattering time $\tau$ in units of $h/\alpha^{2}v_{F}k_{F}$
vs temperature for quasiparticles near the nodal line. In the perturbative
regime, $\tau$ scales logarithmically with temperature and has only
a modest increase per decade of temperature variation compared to
conventional relativistic systems, where $\tau\propto1/T$.}}
\end{figure}

The combination $[\alpha v_{F}]_{(T)}$ does not run, whereas the
ratio $\gamma\equiv v_{z}/v_{F}$ flows toward $1$. Hence, in the
collision-dominated regime $\omega\ll\tau^{-1}$, $\sigma(0,T)$ scales
linearly up to logarithmic corrections, suggesting that the system
behaves as an insulator, as shown in Fig. \ref{fig:conductivity}(a).
In that plot, we use $k_{B}\Lambda_{T}=0.2\times v_{F}k_{F}$, $\gamma=1$
and $\alpha=0.6$. The static charge polarization bubble of a NLSM,
$\Pi(q,0)\sim-N/(2\pi)^{3}k_{F}q/v_{F}$, scales linearly with momentum.
Since the Coulomb interaction $\propto q^{-2}$, the quasiparticles
are partially screened at momenta $q\ll N\alpha k_{F}/2\pi^{2}$,
where interactions decay as $\propto q^{-1}$. Below the cut-off temperature
$\lambda_{T}/\Lambda_{T}\approx N\alpha k_{F}/2\pi^{2}\Lambda\sim0.2$,
the Coulomb interaction is therefore screened by charge polarization
effects, although still long ranged, indicating a crossover in the
$T\to0$ limit. In that regime, the velocity is not further renormalized
by the screened Coulomb interaction and the RG flow stops.

A previous on-shell Wilson-Yukawa RG analysis has indicated the presence
of a screened interacting fixed point in this problem \cite{Huh,Wang}.
In the vicinity of that fixed point, a strong charge renormalization
was found, suggesting a crossover to a Fermi liquid. We point out
that the analysis of Ref. \cite{Huh,Wang} did not incorporate the
nonanalytic structure of the infrared (IR) polarization bubble in
the bosonic propagator, which is relevant in the RG sense. For Dirac
fermions, it has been recently shown \cite{Kruger} that the incorporation
of the IR bubble (which is nonanalytic) in the on-shell propagator
of the bosons is necessary and correctly recovers previous numerical
results based on conformal bootstrap calculations. The fermionic analysis
for NLSMs shown above indicates that the charge is not renormalized
for $T>\Lambda_{T}$ , whereas the velocity is the only physical quantity
that runs in the RG flow in that regime.

From Eq. (\ref{sigma2}) one can extract the scattering time between
collisions, 
\begin{equation}
\tau(T)=0.998\times\frac{\hbar}{\alpha^{2}(T)v_{F}(T)k_{F}}.\label{tau}
\end{equation}
This is the second main result of the paper. In Fermi liquids, the
scattering time diverges as $\tau\propto\hbar\varepsilon_{F}/(k_{B}T)^{2}$,
with $\varepsilon_{F}$ the Fermi energy. Relativistic systems have
a parametrically shorter scattering time ($\tau\sim\hbar/k_{B}T$),
reflecting the absence of screening. The nodal line significantly
enlarges the phase space for collisions among the quasiparticles,
without producing any screening effects at $T\gtrsim\lambda_{T}$.
That further reduces the scattering time, which increases only logarithmically
with decreasing temperature, as shown in Fig. \ref{fig:conductivity}(b).

\section{Shear viscosity}

The shear viscosity $\eta$ is the dissipative response of fluids
to transverse gradients in their velocity field. It is defined after
the strain contribution to the stress tensor away from the equilibrium
distribution \cite{Read} 
\begin{equation}
\delta\langle T_{ij}\rangle=\eta_{ijk\ell}\frac{\partial u_{k}}{\partial x_{\ell}},\label{detaT1}
\end{equation}
where $\mathbf{u}=\partial\boldsymbol{\xi}/\partial t$ is the velocity
field of the fluid, with $\xi_{i}$ being a strain deformation field.
The gradient $u_{ij}\equiv\partial u_{i}/\partial x_{j}=\partial\xi_{ij}/\partial t$
is the time derivative of the strain tensor $\xi_{ij}\equiv\partial\xi_{i}/\partial x_{j}$.
For systems that preserve time-reversal symmetry, the viscosity tensor
is symmetric, obeying the Onsager relation $\eta_{ijk\ell}=\eta_{k\ell ij}$
\cite{Avron}.

The stress tensor can be derived from the change of the Hamiltonian
with respect to the strain tensor, 
\begin{equation}
T_{ij}=\frac{\partial\mathcal{H}}{\partial\xi_{ij}}.\label{eq:}
\end{equation}
In linear response, the first-order contribution of strain to the
Hamiltonian can be shown \cite{Bradlyn} to appear through a term
with the general form 
\begin{equation}
\mathcal{H}_{\xi}=\frac{1}{2}\xi_{ij}(v_{i}k_{j}+k_{j}v_{i}).\label{Hxi}
\end{equation}
From Eq. (\ref{eq:}), the deviation of the expectation value of the
stress tensor $\langle T_{ij}\rangle$ away from equilibrium is 
\begin{equation}
\delta\langle T_{ij}\rangle=N\sum_{\lambda}\int_{\mathbf{k}}(v_{\lambda,\mathbf{k}})_{i}k_{j}\delta f_{\lambda}(\mathbf{k},t),\label{deltaT2}
\end{equation}
from which the shear viscosity in Eq. (\ref{detaT1}) can be extracted.
For details of the derivation, see Appendix D.

Going back to the kinetic equation (\ref{eq:BE}), the second term
on the left gives 
\begin{equation}
-\mathbf{v}_{\lambda,\mathbf{k}}\cdot\nabla_{\mathbf{x}}f_{\lambda}^{0}\left(k\right)=\beta f_{\lambda}^{0}\left(1-f_{\lambda}^{0}\right)\varepsilon_{\lambda,k}^{0}I_{ij}u_{ij}\equiv\Phi_{\lambda,ij}u_{ij},\label{v2}
\end{equation}
with
\begin{equation}
I_{ij}=(v_{\lambda,\mathbf{k}})_{i}k_{j}/\varepsilon_{\lambda,\mathbf{k}}^{0}-(\delta_{ij}/3).\label{I}
\end{equation}
Setting the electric field to zero, the change in the energy spectrum
can be parametrized with the ansatz
\begin{equation}
\varepsilon_{\lambda,k}=\varepsilon_{\lambda,k}^{0}+I_{ij}u_{ij}h_{\lambda}\left(\mathbf{k},t\right),\label{epsilon}
\end{equation}
where $h_{\lambda}(\mathbf{k},t)$ is to be determined by solving
the kinetic equation. Hence, the nonequilibrium correction to the
distribution function due to strain has the form 
\begin{equation}
\delta f_{\lambda}\left(\mathbf{k},t\right)=\beta f_{\lambda}^{0}\left(1-f_{\lambda}^{0}\right)u_{ij}I_{ij}h_{\lambda}\left(\mathbf{k},t\right).\label{deltaf2}
\end{equation}

Defining $\chi_{\lambda,ij}\equiv I_{ij}h_{\lambda}$, the kinetic
equation in the stationary regime ($\omega\to0$) is 
\begin{equation}
\Phi_{\lambda,ij}=\mathcal{C}\chi_{\lambda,ij}.\label{C2}
\end{equation}
The definition of the collision operator follows directly from Eq.
(\ref{C}) under the substitution $\chi_{\lambda,i}\to\chi_{\lambda,ij}$.
In the collinear regime, there are three zero modes that are eigenfunctions
of the collision operator, $\mathcal{C}\chi_{\lambda,ij}=0$, namely,
\begin{equation}
\chi_{\lambda,ij}^{(1)}(\mathbf{k})=\lambda I_{ij},\label{chi1}
\end{equation}
\begin{equation}
\chi_{\lambda,ij}^{(2)}(\mathbf{k})=\varepsilon_{\lambda,k}^{0}I_{ij}\label{chi2}
\end{equation}
and 
\begin{equation}
\chi_{\lambda,ij}^{(3)}(\mathbf{k})=I_{ij}.\label{chi3-1}
\end{equation}
 Those modes correspond to conservation of charge, energy, and number
of particles respectively. The particle number zero mode, however,
does not contribute to the shear viscosity due to particle-hole symmetry
at the nodal line. This mode is orthogonal to the other two and can
be completely decoupled.

Setting a basis with the charge and energy modes $\chi_{\lambda,ij}^{(\alpha)}(\mathbf{k})$,
with $\alpha=1,2$, one can express $\chi_{\lambda,ij}$ as a linear
combination in that basis,
\begin{equation}
\chi_{\lambda,ij}(\mathbf{k})=a_{\beta}\chi_{\lambda,ij}^{(\beta)}(\mathbf{k}).\label{chi}
\end{equation}
If we project the kinetic equation in that basis, namely
\begin{equation}
b^{\alpha}=(\chi_{\lambda,ij}^{(\alpha)},\Phi_{\lambda,ij}),\label{b}
\end{equation}
 and
\begin{equation}
C_{\alpha\beta}=(\chi_{\lambda,ij}^{(\alpha)},\mathcal{C}\chi_{\lambda,ij}^{(\beta)}),\label{c}
\end{equation}
then the solution (\ref{chi}) follows from the determination of the
$a_{\beta}$ coefficients
\begin{equation}
a_{\beta}=b^{\alpha}C_{\alpha\beta}^{-1}.\label{a}
\end{equation}
$C_{\alpha\beta}^{-1}$ is the inverse of a $2\times2$ matrix that
can be evaluated numerically through the momentum integration of the
collision operator in the collinear approximation, as shown in Appendix
C. Substitution in Eqs. (\ref{deltaT2}) and (\ref{deltaf2}) gives
the viscosity tensor
\begin{equation}
\eta_{xixi}(T)=c_{i}(\gamma)N\frac{(k_{B}T)^{3}}{\alpha^{2}v_{F}^{3}(T)},\quad i=y,z\label{eta2}
\end{equation}
where $c_{y}(1)\approx0.569$ and $c_{z}(1)\approx0.759$ for $N=2$.

\subsection{Viscosity-entropy ratio}

The entropy density of a NLSM can be calculated from the entropy of
a noninteracting system dressed by interactions with the renormalized
observables, 
\begin{equation}
s(T)=-k_{B}N\sum_{\lambda}\int_{\mathbf{k}}f_{\lambda}^{0}\ln f_{\lambda}^{0}=\frac{k_{B}^{3}T^{2}k_{F}}{\gamma v_{F}^{2}(T)}\frac{9}{4}\zeta(3),\label{s}
\end{equation}
where $\zeta(3)\approx1.20$ is a zeta function. Allowing the Fermi
velocity and the fine structure constant to be renormalized according
to the RG prescription, the ratio $\eta/s$ is 
\begin{equation}
\frac{\eta}{s}=\frac{\hbar}{k_{B}}\gamma c_{i}(\gamma)\frac{4\zeta(3)}{9}\frac{k_{B}T}{\alpha^{2}(T)v_{F}(T)k_{F}}.\label{eta/s-1}
\end{equation}

\begin{figure}
\begin{centering}
\includegraphics[scale=0.31]{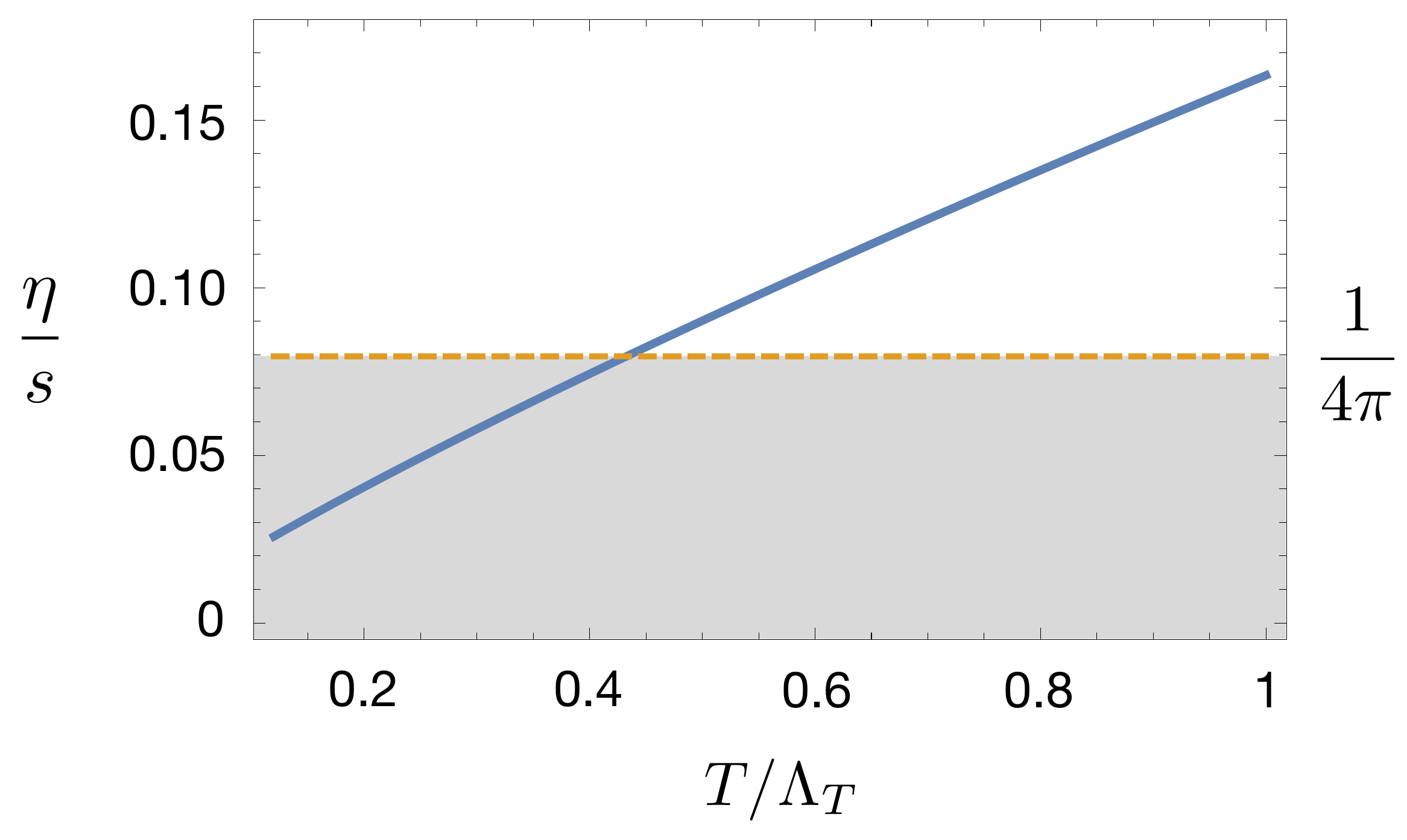}
\par\end{centering}
\caption{{\small{}\label{fig:Ratio}Ratio between the sheer viscosity and the
entropy, $\eta_{xyxy}/s$ (in units of $\hbar/k_{B}$) vs temperature
$T$ normalized by the ultraviolet cutoff $\Lambda_{T}$. We have
set the bare fine structure constant $\alpha=0.6$, $\gamma=1$, and
$k_{B}\Lambda_{T}=0.2\times v_{F}k_{F}$, with $k_{F}$ being the
radius of the nodal line. The horizontal dashed line is the conjectured
lower bound, which is violated in NLSMs at sufficiently low temperature.
At $T\ll\lambda_{T}\sim0.2\Lambda_{T}$, Coulomb interactions are
partially screened by charge polarization effects, suggesting a crossover
(see text).}}
\end{figure}

In Fig. \ref{fig:Ratio}, we plot the temperature dependence of the
shear viscosity-entropy ratio in units of $\hbar/k_{B}$ versus temperature
in units of the temperature cutoff. The horizontal line is the conjectured
lower bound $\eta/s=(1/4\pi)\hbar/k_{B}$. The ratio 
\begin{equation}
\frac{\eta}{s}\propto T\left[1+\frac{\alpha}{4}\ln\left(\frac{\Lambda_{T}}{T}\right)\right]\label{eta/s-2}
\end{equation}
has a quasilinear scaling toward zero with decreasing temperature
$T\in[\lambda_{T},\Lambda_{T}]$, in violation of the lower bound.
The violation reflects the enlarged phase space for collisions at
low temperature in unscreened relativistic systems with a nodal line.
For $T\ll\lambda_{T}$, partial screening effects can lead to the
restoration of a non-universal lower bound, below the one previously
conjectured \cite{Kovtun}.

\section{Discussion}

In the hydrodynamic regime, the usual manifestations of the viscous
flow of electrons in constrained geometries include nonlocal negative
resistance \cite{Baldurin,Crossno,Levitov} and fluid dynamics with
vortex lines \cite{Barenghia}. The very low viscosity compared to
the amount of entropy production, in violation of the conjectured
lower bound, is highly suggestive that NLSMs may exhibit quantum turbulence
\cite{Muller,Shavit,Barenghia}.

In general, observation of hydrodynamics requires quasiparticles with
a relatively short scattering time. Signatures of hydrodynamic behavior
can be detected in the collision-dominated regime through optical
and transport measurements when $k_{B}T\gg\varepsilon_{F},\,\Delta$,
with $\varepsilon_{F}$ being the energy of the Fermi surface and
$\Delta$ being the gap induced by spin-orbit coupling effects or
possible many-body instabilities \cite{Nandkishore,Roy}, including
excitonic phases \cite{Rudenko}. NLSMs that combine inversion, time
reversal, and mirror glide symmetry have nodal lines that are robust
against spin-orbit coupling \cite{Nelson}.

NLSMs are unique in that the nodal line introduces a length scale
that does not generate fully screned interactions, as in Fermi liquids.
That length scale substantially enlarges the size of the phase space
for collision of thermally excited quasiparticles and is responsible
for the unusual temperature scaling of the scattering time in the
hydrodynamic regime. Materials such as ZrSiSe \cite{Shao} have a
large nodal line gapped by a small spin-orbit coupling gap of $\Delta\approx30$
meV, with $v_{F}\Lambda\approx0.4$ eV $(\Lambda_{T}\approx4\times10^{3}K)$
and $v_{F}k_{F}\approx2$eV. In this material, the Fermi velocity
$\hbar v_{F}\approx2$eV$\text{Å}$ is three times smaller than in
graphene. Experimental control over the value of the fine structure
constant can be achieved with experiments on thin films encapsulated
by dielectric materials. In ZrSiSe, for a moderate fine structure
constant $\alpha\approx0.6$ within the perturbative regime, the scattering
length at $T\gg\Delta/k_{B}$,
\begin{equation}
\ell_{s}=v_{F}\tau\sim\frac{\hbar v_{F}}{\alpha^{2}v_{F}k_{F}},\label{ell}
\end{equation}
is of the order of the lattice constant, near the Mott-Ragel-Ioffe
limit, indicating the presence of very strong correlations. We speculate
that hydrodynamic behavior may be observable in a number of different
NLSM materials.

\section{Acknowledgements}

B. U. thanks V. N. Kotov for helpful discussions. The authors acknowledge
the Carl T. Bush Fellowship for partial support. B. U. acknowledges
NSF Grant DMR-2024864 for support.

\appendix

\section{Quantum kinetics in the hydrodynamic regime}

Following the derivation of Kadanoff \citep{Kadanoff}, the Boltzmann
equation has the general form:
\begin{equation}
\left(\frac{\partial}{\partial t}+\mathbf{v}_{\lambda,\mathbf{k}}\cdot\nabla_{\mathbf{x}}+e\mathbf{E}\cdot\nabla_{\mathbf{k}}\right)f_{\lambda}(\mathbf{x},\mathbf{k},t)=\mathcal{I}_{\textrm{col}}[f_{\lambda}],\label{eq:BE-1}
\end{equation}
where $-\nabla_{\mathbf{x}}U_{\textrm{ext}}\left(\boldsymbol{x},t\right)=e\mathbf{E}$
is the external force, $f_{\lambda}\left(\mathbf{x},k,t\right)$ is
the nonequilibrium Fermi distribution, and
\begin{align}
\mathcal{I}_{\textrm{col}}[f_{\lambda}] & =-f_{\lambda}\left(k,t\right)\left(\bar{\Sigma}_{\lambda,\lambda}^{>}\left(k,t\right)\right)_{\omega=\varepsilon_{\lambda}}\nonumber \\
 & +\left(1-f_{\lambda}\left(k,t\right)\right)\left(\bar{\Sigma}_{\lambda,\lambda}^{<}\left(k,t\right)\right)_{\omega=\varepsilon_{\lambda}}\label{eq:a2}
\end{align}
is the collision term, with \begin{widetext}
\begin{align}
\left(\bar{\Sigma}_{\lambda,\lambda}^{>}\right)_{\omega=\varepsilon_{\lambda}} & =\sum_{\lambda_{1}\lambda_{2}\lambda_{3}}\int_{\mathbf{k}_{1}}\int_{\mathbf{k}_{2}}\int_{\mathbf{k}_{3}}\left(2\pi\right)^{4}\delta\left(\mathbf{p}+\mathbf{p}_{1}-\mathbf{p}_{2}-\mathbf{p}_{3}\right)\delta\left(\varepsilon_{\lambda,k}+\varepsilon_{\lambda_{1},k_{1}}-\varepsilon_{\lambda_{2},k_{2}}-\varepsilon_{\lambda_{3},k_{3}}\right)\nonumber \\
 & \times\left[NV\left(\mathbf{k}-\mathbf{k}_{2}\right)V\left(\mathbf{k}-\mathbf{k}_{2}\right)M_{\lambda_{3}\lambda_{1}}M_{\lambda_{1}\lambda_{3}}M_{\lambda\lambda_{2}}M_{\lambda_{2}\lambda}f_{\lambda_{1}}\left(1-f_{\lambda_{2}}\right)\left(1-f_{\lambda_{3}}\right)\right.\nonumber \\
 & \left.-V\left(\mathbf{k}-\mathbf{k_{2}}\right)V\left(\mathbf{k}-\mathbf{k_{3}}\right)M_{\lambda\lambda_{2}}M_{\lambda_{2}\lambda_{1}}M_{\lambda_{1}\lambda_{3}}M_{\lambda_{3}\lambda}f_{\lambda_{1}}\left(1-f_{\lambda_{2}}\right)\left(1-f_{\lambda_{3}}\right)\right]\label{eq:3Dself-1-1}\\
\nonumber \\
\left(\bar{\Sigma}_{\lambda,\lambda}^{<}\right)_{\omega=\varepsilon_{\lambda}} & =\left\{ f\leftrightarrow1-f\right\} .
\end{align}
\end{widetext}$V\left(\mathbf{k}\right)=4\pi e^{2}/\mathbf{k}^{2}$
is the Coulomb interaction and $M$ is a tensor in the quasiparticle-hole
basis. Explicitly,
\begin{equation}
M_{\lambda\lambda_{1}}\left(\mathbf{\mathbf{k}},\mathbf{k}_{1}\right)\equiv\left[U_{\mathbf{\mathbf{k}}}^{-1}U_{\mathbf{k}_{1}}\right]_{\lambda\lambda_{1}},\label{eq:M}
\end{equation}
with $U_{\mathbf{k}}$ being a unitary transformation that diagonalizes
the Hamiltonian.

For nodal-line semimetals (NLSMs),
\begin{align}
\mathcal{H}_{0} & =\frac{k_{r}^{2}-k_{F}^{2}}{2m}\sigma_{x}+v_{z}k_{z}\sigma_{y}\nonumber \\
 & \approx\frac{\left(2k_{F}\right)\left(k_{r}-k_{F}\right)}{2m}\sigma_{x}+v_{z}k_{z}\sigma_{y}\nonumber \\
 & =v_{F}\delta k_{r}\sigma_{x}+v_{z}\delta k_{z}\sigma_{y}\nonumber \\
 & \equiv v_{F}\left(h_{x}\sigma_{x}+h_{y}\sigma_{y}\right),\label{eq:H}
\end{align}
or
\begin{equation}
\mathcal{H}_{0}=\left(\begin{array}{cc}
0 & \mathsf{H}\\
\mathsf{H}^{*} & 0
\end{array}\right),
\end{equation}
where $\mathsf{H}=h_{x}+ih_{y}$,
\[
\left|\mathbf{h}\right|=\left|\mathsf{H}\right|=\sqrt{\left(h_{x}\right)^{2}+\left(h_{y}\right)^{2}}=h.
\]
$\delta\mathbf{k}$ is a relative momentum from the node line. The
Hamiltonian can be diagonalized in the quasi-particle and quasi-hole
basis with their energy $\pm v_{F}h$. We assign each basis as $\lambda=\pm1$,
and thus $\varepsilon_{\lambda,k}=\lambda v_{F}h$ where $\lambda=+1$
corresponds to a excited particle and $\lambda=-1$ to a excited hole.
The unitary transformation matrix is
\begin{equation}
U_{\mathbf{k}}=\frac{1}{\sqrt{2}}\left(\begin{array}{cc}
1 & 1\\
\mathsf{H}/h & -\mathsf{H}/h
\end{array}\right),
\end{equation}
and thus, the tensor $M$ is 
\begin{equation}
M_{\lambda\lambda_{1}}\left(\mathbf{\mathbf{k}},\mathbf{k}_{1}\right)=\frac{1}{2}\left(1+\lambda\lambda_{1}\frac{\mathsf{H}^{*}\mathsf{H}_{1}}{hh_{1}}\right).
\end{equation}
The velocity of quasiparticles in the Boltzmann equation is, by definition,
\begin{align}
\mathbf{v}_{\lambda,\mathbf{k}} & =\frac{\partial\varepsilon_{\lambda,k}}{\partial k_{i}}\nonumber \\
 & =\frac{\lambda v_{F}}{h}\left(h_{x}\frac{\partial h_{x}}{\partial k_{x}},h_{x}\frac{\partial h_{x}}{\partial k_{y}},h_{y}\frac{\partial h_{y}}{\partial k_{z}}\right).\label{eq:v-1}\\
\nonumber 
\end{align}

\subsection{Linearized Boltzmann equation}

Starting from the the nonequilibrium correction of the distribution
function due to the presence of an external electric field, 
\begin{equation}
\delta f_{\lambda}\left(\mathbf{k},\omega\right)=\beta f_{\lambda}^{0}\left(1-f_{\lambda}^{0}\right)e\mathbf{E}\left(\omega\right)\cdot\mathbf{v}_{\lambda,\mathbf{k}}g_{\lambda}(\mathbf{k},\omega).\label{deltaf-1}
\end{equation}
where $\chi_{\lambda,i}\equiv(v_{\lambda,\mathbf{k}})_{i}g_{\lambda}$,
with $g_{\lambda}$ being a function to be determined by solving the
quantum kinetic equation (\ref{eq:BE}). The left-hand side of that
equation is 
\begin{equation}
\left(i\omega\chi_{\lambda}\left(k,\omega\right)-1\right)\beta eE_{i}\left(\omega\right)\left(\mathbf{v}_{\mathbf{k}}\right)_{i}f_{\lambda}^{0}\left(1-f_{\lambda}^{0}\right).\label{eq:LHS}
\end{equation}
Defining $f_{\lambda_{i}}\equiv f_{\lambda_{i}}\left(k_{i}\right)$,
the collision term in the right-hand side is \begin{widetext}
\begin{align}
\mathcal{I}_{\text{col}}[f_{\lambda}] & =\sum_{\lambda_{1}\lambda_{2}\lambda_{3}}\int_{\mathbf{k}_{1}}\int_{\mathbf{k}_{2}}\int_{\mathbf{k}_{3}}\left(2\pi\right)^{4}\delta^{3}\left(\mathbf{k}+\mathbf{k}_{1}-\mathbf{k}_{2}-\mathbf{k}_{3}\right)\delta\left(\varepsilon_{\lambda}+\varepsilon_{\lambda_{1}}-\varepsilon_{\lambda_{2}}-\varepsilon_{\lambda_{3}}\right)\nonumber \\
 & \times\left[NV\left(\mathbf{k}-\mathbf{k}_{2}\right)^{2}W_{\lambda\lambda_{1}\lambda_{2}\lambda_{3}}-V\left(\mathbf{k}-\mathbf{k_{2}}\right)V\left(\mathbf{k}-\mathbf{k_{3}}\right)Y_{\lambda\lambda_{1}\lambda_{2}\lambda_{3}}\right]\nonumber \\
 & \times\left[\left(1-f_{\lambda}\right)\left(1-f_{\lambda_{1}}\right)f_{\lambda_{2}}f_{\lambda_{3}}-f_{\lambda}f_{\lambda_{1}}\left(1-f_{\lambda_{2}}\right)\left(1-f_{\lambda_{3}}\right)\right],\label{eq:RHS}
\end{align}
with $N$ being the fermionic degeneracy, and 
\begin{align}
W_{\lambda\lambda_{1}\lambda_{2}\lambda_{3}} & =M_{\lambda\lambda_{2}}M_{\lambda_{2}\lambda}M_{\lambda_{3}\lambda_{1}}M_{\lambda_{1}\lambda_{3}}\nonumber \\
Y_{\lambda\lambda_{1}\lambda_{2}\lambda_{3}} & =M_{\lambda\lambda_{2}}M_{\lambda_{1}\lambda_{3}}M_{\lambda_{3}\lambda}M_{\lambda_{2}\lambda_{1}},\label{WQ}
\end{align}
where $M_{\lambda\lambda_{1}}\equiv M_{\lambda\lambda_{1}}(\mathbf{k},\mathbf{k}_{1})$,
and so on.

The third line of (\ref{eq:RHS}) has two terms with four $f$ functions.
One should expand it in eight terms to linear order in $\delta f$,
with three $f^{0}$ and one $\delta f$. We can simplify them using
\begin{align}
f_{-\lambda}^{0}f_{-\lambda_{1}}^{0}f_{\lambda_{2}}^{0}f_{\lambda_{3}}^{0} & =e^{\left(\lambda v_{r}k^{\prime}+\lambda v_{r}k_{1}^{\prime}\right)\beta}f_{\lambda}^{0}f_{\lambda_{1}}^{0}f_{\lambda_{2}}^{0}f_{\lambda_{3}}^{0}\nonumber \\
 & =e^{\left(\lambda v_{r}k_{2}^{\prime}+\lambda v_{r}k_{3}^{\prime}\right)\beta}f_{\lambda}^{0}f_{\lambda_{1}}^{0}f_{\lambda_{2}}^{0}f_{\lambda_{3}}^{0}\nonumber \\
 & =f_{\lambda}^{0}f_{\lambda_{1}}^{0}f_{-\lambda_{2}}^{0}f_{-\lambda_{3}}^{0}.
\end{align}
 which is restricted by the energy conservation. After some straightforward
algebra, we find 
\begin{align}
\mathcal{I}_{\text{col}}[f_{\lambda}]= & -\sum_{\lambda_{1}\lambda_{2}\lambda_{3}}\int_{\mathbf{k}_{1}}\int_{\mathbf{k}_{2}}\int_{\mathbf{k}_{3}}\left(2\pi\right)^{4}\delta^{3}\left(\mathbf{k}+\mathbf{k}_{1}-\mathbf{k}_{2}-\mathbf{k}_{3}\right)\delta\left(\varepsilon_{\lambda}+\varepsilon_{\lambda_{1}}-\varepsilon_{\lambda_{2}}-\varepsilon_{\lambda_{3}}\right)\nonumber \\
 & \times\left[NV^{2}\left(\mathbf{k}-\mathbf{k}_{2}\right)W_{\lambda\lambda_{1}\lambda_{2}\lambda_{3}}-V\left(\mathbf{k}-\mathbf{k}_{2}\right)V\left(\mathbf{k}-\mathbf{k}_{3}\right)Y_{\lambda\lambda_{1}\lambda_{2}\lambda_{3}}\right]\beta eE_{i}\left(\omega\right)f_{\lambda}^{0}f_{\lambda_{1}}^{0}f_{-\lambda_{2}}^{0}f_{-\lambda_{3}}^{0}\nonumber \\
 & \times\left[\chi_{i}\left(\lambda,k\right)+\chi_{i}\left(\lambda_{1},k_{1}\right)-\chi_{i}\left(\lambda_{2},k_{2}\right)-\chi_{i}\left(\lambda_{3},k_{3}\right)\right],\label{eq:RHS2}
\end{align}
with the collision matrix element 
\begin{equation}
\mathcal{M}_{\lambda\lambda_{1}\lambda_{2}\lambda_{3}}^{\text{Col}}=NV^{2}\left(\mathbf{k}-\mathbf{k}_{2}\right)W_{\lambda\lambda_{1}\lambda_{2}\lambda_{3}}-V\left(\mathbf{k}-\mathbf{k}_{2}\right)V\left(\mathbf{k}-\mathbf{k}_{3}\right)Y_{\lambda\lambda_{1}\lambda_{2}\lambda_{3}}.\label{M2}
\end{equation}

Defining 
\begin{equation}
\phi_{\lambda,i}(k)\equiv\beta f_{\lambda}^{0}(1-f_{\lambda}^{0})(v_{\lambda,\mathbf{k}})_{i},\label{eq:phi}
\end{equation}
equating the left- and the right-hand side of the quantum Boltzmann
equation, Eq. (\ref{eq:LHS}) and (\ref{eq:RHS2}), we have 
\begin{equation}
\phi_{\lambda,i}=\mathcal{C}\chi_{\lambda,i}+i\omega g_{\lambda}\phi_{\lambda,i},\label{phiEq}
\end{equation}
with $\mathcal{C}$ being the collision operator as defined in the
main text,
\begin{align}
\mathcal{C}\chi_{\lambda,i} & =\sum_{\lambda_{1}\lambda_{2}\lambda_{3}}\int_{\mathbf{k}_{1}}\int_{\mathbf{k}_{2}}\int_{\mathbf{k}_{3}}\left(2\pi\right)^{4}\delta^{3}\left(\mathbf{k}+\mathbf{k}_{1}-\mathbf{k}_{2}-\mathbf{k}_{3}\right)\delta\left(\varepsilon_{\lambda,k}^{0}+\varepsilon_{\lambda_{1},k_{1}}^{0}-\varepsilon_{\lambda_{2},k_{2}}^{0}-\varepsilon_{\lambda_{3},k_{3}}^{0}\right)\mathcal{M}_{\lambda\lambda_{1}\lambda_{2}\lambda_{3}}^{\textrm{Col}}\nonumber \\
 & \quad\qquad\qquad\times f_{\lambda}^{0}f_{\lambda_{1}}^{0}f_{-\lambda_{2}}^{0}f_{-\lambda_{3}}^{0}\left[\chi_{\lambda,i}\left(k\right)+\chi_{\lambda_{1},i}\left(k_{1}\right)-\chi_{\lambda_{2},i}\left(k_{2}\right)-\chi_{\lambda_{3},i}\left(k_{3}\right)\right].\label{C-1}
\end{align}

\section{Collinear approximation}

\subsection{Collision phase space}

Due to the Coulomb potential $V\left(\mathbf{k}-\mathbf{k}_{2}\right)$
and $V\left(\mathbf{k}-\mathbf{k}_{3}\right)$ in the integrand of
the collision operator, the integral is governed by small momentum
transfer due to collision processes. In the collinear approximation,
where the four momenta are nearly aligned to each other around the
nodal line, we can define the momenta
\begin{align}
\mathbf{k} & =\left(k_{r},0,k_{z}\right)=\left(\delta k_{r}+k_{F},0,k_{z}\right)\\
\mathbf{k}_{1} & \approx\left(k_{1r},k_{1\perp},k_{1z}\right)=\left(\delta k_{1r}+k_{F},k_{1\perp},k_{1z}\right)\\
\mathbf{k}_{2} & \approx\left(k_{2r},k_{2\perp},k_{2z}\right)=\left(\delta k_{2r}+k_{F},k_{2\perp},k_{2z}\right)\\
\mathbf{k}_{3} & \approx\left(k_{3r},k_{3\perp},k_{3z}\right)=\left(\delta k_{3r}+k_{F},k_{3\perp},k_{3z}\right),
\end{align}
where we assume that the $\perp$ components are small compared to
the radius of the nodal line $k_{F}$. The phase space for collision
processes is set by conservation of energy,
\[
\delta\left(\varepsilon_{\lambda,\mathbf{k}}^{0}+\varepsilon_{\lambda_{1},\mathbf{k}_{1}}^{0}-\varepsilon_{\lambda_{2},\mathbf{k}_{2}}^{0}-\varepsilon_{\lambda_{3},\mathbf{k}_{3}}^{0}\right).
\]
We now expressing it in terms of the dimensionless variables,
\begin{align}
x\equiv v_{F}\beta(\delta k_{r}) & ,\qquad y\equiv v_{z}\beta k_{z},\qquad\kappa_{0}\equiv v_{F}k_{F}\beta,\qquad r^{2}\equiv x^{2}+y^{2},\label{eq:dim1}
\end{align}
and
\begin{align}
x_{n}\equiv v_{F}\beta(\delta k_{nr}) & ,\qquad y_{n}\equiv v_{z}\beta k_{nz},\qquad\xi_{n}\equiv v_{F}\beta k_{n\perp},\qquad r_{n}^{2}\equiv x_{n}^{2}+y_{n}^{2},\label{dim2}
\end{align}
with $n=1,2,3$. Performing a suitable change of variables $\mathbf{k}_{2}\to\mathbf{k}-\mathbf{k}_{2}$
and $\mathbf{k}_{3}\to\mathbf{k}_{1}-\mathbf{k}_{3}$,
\begin{equation}
\beta\delta(D)\equiv\delta\left(\varepsilon_{\lambda,\mathbf{k}}^{0}+\varepsilon_{\lambda_{1},\mathbf{k}_{1}}^{0}-\varepsilon_{\lambda_{2},\mathbf{k}+\mathbf{k}_{2}}^{0}-\varepsilon_{\lambda_{3},\mathbf{k}_{1}-\mathbf{k}_{3}}^{0}\right),\label{deltaF}
\end{equation}
where 
\begin{align}
D & =\lambda r+\lambda_{1}\sqrt{\left(x_{1}+\frac{\xi_{1}^{2}}{2\kappa_{0}}\right)^{2}+y_{1}^{2}}-\lambda_{2}\sqrt{\left(x+x_{2}+\frac{\xi_{2}^{2}}{2\kappa_{0}}\right)^{2}+\left(y+y_{2}\right)^{2}}-\lambda_{3}\sqrt{\left(x_{1}-x_{2}+\frac{\left(\xi_{1}-\xi_{2}\right)^{2}}{2\kappa_{0}}\right)^{2}+\left(y_{1}-y_{2}\right)},\label{D}
\end{align}
while at the same time
\begin{align}
V\left(\mathbf{k}-\mathbf{k}_{2}\right)\longrightarrow\bar{V_{1}} & =\frac{1}{\left(x_{2}\right)^{2}+\gamma^{-2}\left(y_{2}\right)^{2}+\left(\xi_{2}\right)^{2}}\label{V1}\\
V\left(\mathbf{k}-\mathbf{k}_{3}\right)\longrightarrow\bar{V_{2}} & =\left(\frac{1}{\left(x-x_{1}+x_{2}\right)^{2}+\gamma^{-2}\left(y-y+y_{2}\right)^{2}+\left(\xi_{1}-\xi_{2}\right)^{2}}\right),\label{V2}
\end{align}
after using momentum conservation $\mathbf{k}+\mathbf{k}_{1}-\mathbf{k}_{2}-\mathbf{k}_{3}=0$.

Since $\xi_{1}$ and $\xi_{2}$ are much smaller than $\kappa_{0}$,
we can rewrite the argument of the $\delta$ function $D$ as
\begin{align}
D & \approx\bar{A}+\lambda_{1}\frac{x_{1}\xi_{1}^{2}}{2r_{1}\kappa_{0}}-\lambda_{2}\frac{\left(x+x_{2}\right)\xi_{2}^{2}}{2\left|\mathbf{r}+\mathbf{r}_{2}\right|\kappa_{0}}-\lambda_{3}\frac{\left(x_{1}-x_{2}\right)\left(\xi_{1}-\xi_{2}\right)^{2}}{2\left|\mathbf{r}_{1}-\mathbf{r}_{2}\right|\kappa_{0}}\nonumber \\
 & =-\left(\frac{\lambda_{3}\left(x_{1}-x_{2}\right)}{2\left|\mathbf{r}_{1}-\mathbf{r}_{2}\right|\kappa_{0}}+\frac{\lambda_{2}\left(x+x_{2}\right)}{2\left|\mathbf{r}+\mathbf{r}_{2}\right|\kappa_{0}}\right)\xi_{2}^{2}+\frac{\lambda_{3}\left(x_{1}-x_{2}\right)}{\left|\mathbf{r}_{1}-\mathbf{r}_{2}\right|\kappa_{0}}\xi_{1}\xi_{2}-\left(\frac{\lambda_{3}\left(x_{1}-x_{2}\right)}{2\left|\mathbf{r}_{1}-\mathbf{r}_{2}\right|\kappa_{0}}-\frac{\lambda_{1}x_{1}}{2r_{1}\kappa_{0}}\right)\xi_{1}^{2}+\bar{A}\nonumber \\
 & \equiv-\frac{w_{1}}{2\kappa_{0}}\left(\xi_{1}^{2}-2\frac{w_{2}}{w_{1}}\xi_{2}\xi_{1}+\frac{w_{3}}{w_{1}}\xi_{2}^{2}-\frac{\bar{A}}{w_{1}}\right)\label{eq:D2}
\end{align}
\end{widetext}where
\begin{align}
\bar{A} & \equiv\lambda r+\lambda_{1}r_{1}-\lambda_{2}\left|\mathbf{r}+\mathbf{r}_{2}\right|-\lambda_{3}\left|\mathbf{r}_{1}-\mathbf{r}_{2}\right|\\
\left|\mathbf{r}+\mathbf{r}_{2}\right| & \equiv\sqrt{\left(x+x_{2}\right)^{2}+\left(y+y_{2}\right)^{2}}\\
\left|\mathbf{r}_{1}-\mathbf{r}_{2}\right| & \equiv\sqrt{\left(x_{1}-x_{2}\right)^{2}+\left(y_{1}-y_{2}\right)^{2}},
\end{align}
and $w_{i}$ ($i=1,2,3$) are functions of the dimensionles variables
$x,y,x_{i},y_{i}$. $D$ is a quadratic function of $\xi_{1}$. We
can then express the $\delta$ function as
\begin{equation}
\delta\left(\bar{D}\left(\xi_{1}\right)\right)=\sum_{i=\pm}\frac{\delta\left(\xi_{1}-\xi_{i}\right)}{\left|D^{\prime}\left(\xi_{i}\right)\right|},
\end{equation}
where $D^{\prime}$ is the first derivative of $D$, and $\xi_{i=\pm}$
are the two roots of the quadratic function, namely
\begin{equation}
\xi_{\pm}=\frac{w_{2}}{w_{1}}\xi_{2}\pm\sqrt{\left(\frac{w_{2}}{w_{1}}\xi_{2}\right)^{2}-\left(\frac{w_{3}}{w_{1}}\xi_{2}^{2}-\frac{\bar{A}}{w_{1}}\right)}.
\end{equation}
Hence,
\begin{align}
\left|\bar{D}^{\prime}\left(\xi_{i}\right)\right| & =\left|\frac{w_{1}}{2\kappa_{0}}\left[\left(2\xi_{1}-\xi_{+}-\xi_{-}\right)\right]_{\xi_{1}=\xi_{i}}\right|\nonumber \\
 & =\left|\frac{w_{1}}{2\kappa_{0}}\left(\xi_{+}-\xi_{-}\right)\right|\nonumber \\
 & =\left|\frac{w_{1}}{2\kappa_{0}}\sqrt{\left(\frac{w_{2}}{w_{1}}\xi_{2}\right)^{2}-\left(\frac{w_{3}}{w_{1}}\xi_{2}^{2}-\frac{\bar{A}}{w_{1}}\right)}\right|.
\end{align}
Thus,
\begin{align}
\delta\left(D\left(\xi_{1}\right)\right) & =\kappa_{0}\left|\frac{1}{\sqrt{\left(w_{1}w_{2}\xi_{2}\right)^{2}-\left(w_{1}w_{3}\xi_{2}^{2}-w_{1}\bar{A}\right)}}\right|\label{delta3}\\
 & \quad\times\left(\delta\left(\xi_{1}-\xi_{1+}\right)+\delta\left(\xi_{1}-\xi_{1-}\right)\right)
\end{align}
It is clear that the phase space has a logarithmic divergence in the
$\xi_{2}$ variable when $\bar{A}\rightarrow0$. At the same time,
the Coulomb interaction terms $\bar{V}_{1}$ and $\bar{V}_{2}$ defined
in Eqs. (\ref{V1}) and (\ref{V2}) decay quickly to zero with $\xi_{2}$
when it is large. Thus, there are two important regions of the integrand
in phase space: $\bar{A}\rightarrow0$ and $\xi_{2}\rightarrow0$.
This phase space argument justifies the validity of the collinear
approximation, with which the conductivity and the shear viscosity
were calculated.

\subsubsection{Calculation of the conductivity}

The variational functional of the conductivity is
\begin{equation}
Q\left[a^{(e)}\right]\equiv\left(\chi_{\lambda,i},\phi_{\lambda,i}\right)-\frac{1}{2}\left(\chi_{\lambda,i},C\chi_{\lambda,i}+i\omega a^{(e)}\phi_{\lambda,i}\right).\label{eq:Q-1}
\end{equation}
We define the inner product $(a_{\lambda,i},b_{\lambda,i})\equiv\sum_{\lambda,i}\int_{\mathbf{k}}a_{\lambda,i}(\mathbf{k})b_{\lambda,i}(\mathbf{k})$,
with
\[
\frac{\partial Q}{\partial a^{(e)}}=0,
\]
with $a^{(e)}(\omega)$ being the variational function corresponding
to charge conservation in the collinear regime. For convenience, after
multiplying the factor $v_{F}\beta^{3}$ in both sides of Eq. (\ref{eq:Q-1}),
the first term is \begin{widetext}
\begin{align}
v_{z}\beta^{3}\left(\chi_{\lambda,i},\phi_{\lambda,i}\right) & =a^{(e)}v_{F}\beta^{3}\sum_{\lambda}\int_{\mathbf{k}}\lambda^{2}v_{r}^{2}\int\frac{d^{3}k}{\left(2\pi\right)^{3}}\frac{1}{\left(e^{\lambda v_{r}h\beta}+1\right)\left(e^{-\lambda v_{r}h\beta}+1\right)}\nonumber \\
 & =a^{(e)}\frac{v_{F}k_{F}\beta}{\pi}\left(\gamma^{-1}+\gamma\right)\int dr\frac{r}{\left(e^{r}+1\right)\left(e^{-r}+1\right)}\nonumber \\
 & =a^{(e)}\kappa_{0}\frac{\ln\left(2\right)}{2\pi}\left(\gamma^{-1}+\gamma\right),
\end{align}
where $\gamma\equiv v_{z}/v_{F}$, and
\begin{equation}
\int\frac{d^{3}k}{\left(2\pi\right)^{3}}\rightarrow\int k_{F}\frac{dk_{r}\,dk_{z}}{\left(2\pi\right)^{3}}d\phi\rightarrow k_{F}\int\frac{d\delta k_{r}\,d\delta k_{z}}{\left(2\pi\right)^{3}}d\phi\rightarrow\frac{k_{F}}{v_{r}v_{z}\beta^{2}}\int\frac{dx\,dy}{\left(2\pi\right)^{3}}d\phi.
\end{equation}

To calculate the second term, we consider the dominant processes in
the near collinear regime, which conserve the number of particles
and holes. We have
\begin{align}
\frac{v_{F}\beta^{3}}{2}\left(\chi_{\lambda,i},\mathcal{C}\chi_{\lambda,i}+i\omega a^{(e)}\phi_{\lambda,i}\right) & =\frac{v_{F}\beta^{3}}{8}\sum_{\lambda_{i}}\int\frac{d^{3}k}{\left(2\pi\right)^{3}}\frac{d\delta k_{1r}d\delta k_{1z}dk_{1\perp}}{\left(2\pi\right)^{3}}\frac{d\delta k_{2r}d\delta k_{2r}dk_{2\perp}}{\left(2\pi\right)^{3}}\frac{d\delta k_{3r}d\delta_{3r}dk_{3\perp}}{\left(2\pi\right)^{3}}\nonumber \\
 & \qquad\times2\pi\delta\left(\lambda v_{F}h+\lambda_{1}v_{F}h_{1}-\lambda_{2}v_{F}h_{2}-\lambda_{3}v_{F}h_{3}\right)\left(2\pi\right)^{3}\delta^{3}\left(\mathbf{k}+\mathbf{k}_{1}-\mathbf{k}_{2}-\mathbf{k}_{3}\right)\nonumber \\
 & \qquad\times f_{\lambda}^{0}f_{\lambda_{1}}^{0}f_{-\lambda_{2}}^{0}f_{-\lambda_{3}}^{0}\left[NV^{2}\left(\mathbf{k}-\mathbf{k}_{2}\right)W_{\lambda\lambda_{1}\lambda_{2}\lambda_{3}}-V\left(\mathbf{k}-\mathbf{k}_{2}\right)V\left(\mathbf{k}-\mathbf{k}_{3}\right)Y_{\lambda\lambda_{1}\lambda_{2}\lambda_{3}}\right]\nonumber \\
 & \qquad\times\left(a^{(e)}\right)^{2}\left[\mathbf{v}_{\lambda,\mathbf{k}}+\mathbf{v}_{\lambda_{1},\mathbf{k}_{1}}-\mathbf{v}_{\lambda_{2},\mathbf{k}_{2}}-\mathbf{v}_{\lambda_{3},\mathbf{k}_{3}}\right]^{2}+\frac{i\omega}{2}\left(a^{(e)}\right)^{2}\kappa_{0}\frac{\ln\left(2\right)}{\pi}\left(\gamma^{-1}+\gamma\right)\nonumber \\
 & \equiv\frac{\kappa_{0}^{2}\alpha^{2}}{\beta}\left[a^{(e)}\right]^{2}I(\gamma)+i\omega\left[a^{(e)}\right]^{2}\kappa_{0}\frac{\ln\left(2\right)}{2\pi}\left(\gamma^{-1}+\gamma\right),\label{I-1}
\end{align}
where $\alpha\equiv e^{2}/v_{F}$, and $I(\gamma$) is a dimensionless
number. The extra factor of $\frac{1}{4}$ on the right-hand side
is due to the symmetrization in the four momenta. In terms of the
dimensionless variables (\ref{eq:dim1}) and (\ref{dim2}), the combination
$\left[a^{(e)}\right]^{2}I(\gamma)$ can be written in the collinear
approximation as
\begin{align}
\left[a^{(e)}\right]^{2}I(\gamma) & \approx-\frac{1}{8\gamma^{3}}\left(4\pi\right)^{2}\int\frac{dx\,dy}{2\pi^{2}}\frac{dx_{1}dy_{1}d\xi_{1}}{\left(2\pi\right)^{3}}\frac{dx_{2}dy_{2}d\xi_{2}}{\left(2\pi\right)^{3}}2\pi\delta\left(D\right)f_{\lambda}^{0}f_{\lambda_{1}}^{0}f_{-\lambda_{2}}^{0}f_{-\lambda_{3}}^{0}\nonumber \\
 & \qquad\quad\times\left(N\bar{V}_{1}^{2}W_{\lambda\lambda_{1}\lambda_{2}\lambda_{3}}-\bar{V_{1}}\bar{V}_{2}Y_{\lambda\lambda_{1}\lambda_{2}\lambda_{3}}\right)\left(a^{(e)}\right)^{2}\left[\bar{X}_{\lambda\lambda_{1}\lambda_{2}\lambda_{3}}\right]^{2},\label{AI}
\end{align}
where
\begin{equation}
\left(\bar{X}_{\lambda\lambda_{1}\lambda_{2}\lambda_{3}}\right)^{2}\equiv\left(\lambda\frac{x}{r}+\lambda_{1}\frac{x_{1}}{r_{1}}-\lambda_{2}\frac{x+x_{2}}{\left|\mathbf{r}+\mathbf{r}_{2}\right|}-\lambda_{3}\frac{x_{1}-x_{2}}{\left|\mathbf{r}_{1}-\mathbf{r}_{2}\right|}\right)^{2}+\left(\lambda\frac{y}{r}+\lambda_{1}\frac{y_{1}}{r_{1}}-\lambda_{2}\frac{y+y_{2}}{\left|\mathbf{r}+\mathbf{r}_{2}\right|}-\lambda_{3}\frac{y_{1}-y_{2}}{\left|\mathbf{r}_{1}-\mathbf{r}_{2}\right|}\right)^{2}.\label{eq:Xbar2}
\end{equation}
$D$, $\bar{V}_{1}$, and $\bar{V}_{2}$ are given in Eq. (\ref{D}),
(\ref{V1}), and (\ref{V2}). The $W$ and $Y$ tensors follow from
Eqs. (\ref{eq:M}) and (\ref{WQ}) with the substitution $h_{x}\rightarrow x$,
$h_{y}\rightarrow y$ and so on. The integral is performed enforcing
the restriction in momentum space $\left(w_{1}w_{2}\xi_{2}\right)^{2}-\left(w_{1}w_{3}\xi_{2}^{2}-w_{1}\bar{A}\right)>0$
after integrating $\xi_{1}$ out through the $\delta$ function (\ref{delta3}).
From Eq. (\ref{I-1}),
\begin{align}
\frac{\partial Q}{\partial a^{(e)}} & =-\kappa_{0}\frac{\ln\left(2\right)}{2\pi}\left(\gamma^{-1}+\gamma\right)+\frac{\kappa_{0}^{2}\alpha^{2}}{\beta}a^{(e)}I\left(\gamma\right)+i\omega a^{(e)}\kappa_{0}\frac{\ln\left(2\right)}{2\pi}\left(\gamma^{-1}+\gamma\right)=0.
\end{align}
This implies that
\begin{equation}
a^{(e)}(\omega)=\frac{\beta}{\kappa_{0}\alpha^{2}c\left(\gamma\right)+i\omega\beta},
\end{equation}
where
\begin{equation}
c\left(\gamma\right)=\frac{2\pi}{\ln2\left(\gamma^{-1}+\gamma\right)}I\left(\gamma\right)\text{ }
\end{equation}
In the near collinear approximation, we find $c\left(\gamma=1\right)\approx1.034$
for $N=2$. When the nodal line is spin polarized, with $N=1$, $c(1)\approx0.361$.
In the two anisotropic limits $\gamma\to0$ and $\gamma\to\infty$,
$c\left(\gamma\right)$ is proportional to $\gamma^{2}$ and $\gamma^{-2}$
respectively, and scales toward zero.

The conductivity is
\begin{align}
\sigma_{yy}=\sigma_{xx}=\frac{\partial J_{x}}{\partial E_{x}} & =\frac{e^{2}}{\hbar}\sum_{\lambda}\int\frac{d^{3}k}{\left(2\pi\right)^{3}}\left(\mathbf{v}_{\lambda,\mathbf{k}}\right)_{x}\left(\mathbf{v}_{\mathbf{k}}\right)_{x}\beta f_{\lambda}^{(0)}\left(1-f_{\lambda}^{(0)}\right)a^{(e)}\nonumber \\
 & =2\pi\frac{e^{2}}{h}\frac{1}{\kappa_{0}\alpha^{2}c\left(\gamma\right)+i\omega\beta}\frac{2Nk_{F}}{\gamma}\int\frac{1}{2\pi}\cos^{2}\phi d\phi\int\frac{dx}{2\pi}\frac{dy}{2\pi}\frac{x^{2}e^{r}}{r^{2}\left(e^{r}+1\right)^{2}}\nonumber \\
 & =\frac{e^{2}}{2h}k_{F}\frac{1}{\gamma\beta}\frac{N\ln\left(2\right)}{v_{F}k_{F}\alpha^{2}c\left(\gamma\right)+i\omega}\\
 & =\frac{1}{\gamma^{2}}\sigma_{zz}.
\end{align}
\end{widetext}

\subsection{Calculation of the viscosity}

In the collinear regime, we set a basis with the zero modes reflecting
conservation of energy and number of particles $\{\chi_{\lambda,ij}^{(1)},\chi_{\lambda,ij}^{(2)}\}$,
\begin{equation}
\chi_{\lambda,ij}^{(1)}(\mathbf{k})=\lambda I_{ij},\qquad\chi_{\lambda,ij}^{(2)}(\mathbf{k})=\beta\varepsilon_{\lambda,\mathbf{k}}^{0}I_{ij},\label{basis}
\end{equation}
with 
\[
I_{ij}=\sqrt{\frac{3}{2}}\left[(\mathbf{v}_{\lambda,\mathbf{k}})_{i}k_{j}/\varepsilon_{\lambda,\mathbf{k}}^{0}-(\delta_{ij}/3)\right],
\]
as described in the main text. One can express $\chi_{\lambda,ij}$
as a linear combination in that basis. Projecting $b^{\alpha}=(\chi_{\lambda,ij}^{(\alpha)},\Phi_{\lambda,ij})$,
where 
\begin{equation}
\Phi_{\lambda,ij}=\beta f_{\lambda}^{0}\left(1-f_{\lambda}^{0}\right)\varepsilon_{\lambda,k}^{0}I_{ij}
\end{equation}
and $C_{\alpha\beta}=(\chi_{\lambda,ij}^{(\alpha)},\mathcal{C}\chi_{\lambda,ij}^{(\beta)})$,
with $\alpha=1,2$, then the solution of the kinetic equation is 
\begin{equation}
\chi_{\lambda,ij}(\mathbf{k})=a_{\beta}\chi_{\lambda,ij}^{(\beta)}(\mathbf{k})=b^{\alpha}C_{\alpha\beta}^{-1}\chi_{\lambda,ij}^{(\beta)}(\mathbf{k}),\label{chi3}
\end{equation}
where $C_{\alpha\beta}^{-1}$ is the inverse of a $2\times2$ matrix,
and 
\[
a_{\beta}=b^{\alpha}C_{\alpha\beta}^{-1}.
\]

To be specific, one can define two different variational functions
$Q$ with the two modes as 
\begin{align}
Q\left[\chi_{\lambda,ij}^{(1)}\right] & \equiv\left(\chi_{\lambda,ij}^{(1)},\Phi_{\lambda,ij}\right)-\frac{1}{2}\left(\chi_{\lambda,ij}^{(1)},\mathcal{C}\chi_{\lambda,ij}\right)\\
Q\left[\chi_{\lambda,ij}^{(2)}\right] & \equiv\left(\chi_{\lambda,ij}^{(2)},\Phi_{\lambda,ij}\right)-\frac{1}{2}\left(\chi_{\lambda,ij}^{(2)},C\chi_{\lambda,ij}\right).
\end{align}
Minimization results in two equations with the form 
\begin{equation}
\left(\begin{array}{c}
b^{1}\\
b^{2}
\end{array}\right)=\left(\begin{array}{cc}
C_{11} & C_{12}\\
C_{21} & C_{22}
\end{array}\right)\left(\begin{array}{c}
a_{1}\\
a_{2}
\end{array}\right),
\end{equation}
where \begin{widetext}

\begin{align}
C_{\alpha\beta} & =\frac{1}{8}\sum_{\lambda_{i}}^{i=1,2,3}\int_{\mathbf{k}}\int_{\mathbf{k}_{1}}\int_{\mathbf{k}_{2}}\int_{\mathbf{k}_{3}}\left(2\pi\right)^{4}\delta^{3}\left(\mathbf{k}+\mathbf{k}_{1}-\mathbf{k}_{2}-\mathbf{k}_{3}\right)\delta\left(\lambda v_{r}k^{\prime}+\lambda_{1}v_{r}k_{1}^{\prime}-\lambda_{2}v_{r}k_{2}^{\prime}-\lambda_{3}v_{r}k_{3}^{\prime}\right)\mathcal{M}_{\lambda\lambda_{1}\lambda_{2}\lambda_{3}}^{\textrm{Col}}\nonumber \\
 & \times f_{\lambda}^{0}f_{\lambda_{1}}^{0}f_{-\lambda_{2}}^{0}f_{-\lambda_{3}}^{0}\left(\chi_{\lambda,ij}^{(\alpha)}(\mathbf{k})+\chi_{\lambda,ij}^{(\alpha)}(\mathbf{k}_{1})-\chi_{\lambda,ij}^{(\alpha)}(\mathbf{k}_{2})-\chi_{\lambda,ij}^{(\alpha)}(\mathbf{k}_{3})\right)\left(\chi_{\lambda,ij}^{(\beta)}(\mathbf{k})+\chi_{\lambda,ij}^{(\beta)}(\mathbf{k}_{1})-\chi_{\lambda,ij}^{(\beta)}(\mathbf{k}_{2})-\chi_{\lambda,ij}^{(\beta)}(\mathbf{k}_{3})\right).\label{eq:Cmunu-1}
\end{align}
and 
\begin{align*}
b^{\alpha} & =\sum_{\lambda}\int_{\mathbf{k}}f_{\lambda}^{0}f_{-\lambda}^{0}\lambda\beta\varepsilon_{\lambda,k}^{0}I_{ij}\left(\mathbf{k}\right)\chi_{\lambda,ij}^{(\alpha)}(\mathbf{k})=\kappa_{0}\times\begin{cases}
\frac{\pi}{12}\left(\frac{3}{2\gamma}-1+\frac{3\gamma}{2}\right) & (\alpha=1)\\
\frac{9}{4\pi}\zeta(3)\left(\frac{3}{2\gamma}-1+\frac{3\gamma}{2}\right). & (\alpha=2)
\end{cases}.
\end{align*}
We calculate the $C_{\alpha\beta}$ matrix numerically in the near
collinear approximation. Inverting the resulting matrix, the coefficients
$a_{\alpha}$ ($\alpha=1,2)$ for $N=2$ are 
\begin{equation}
b^{\alpha}C_{\alpha\beta}^{-1}=\left(a_{1},a_{2}\right)\approx\frac{1}{\kappa_{0}\alpha^{2}}\left(-1.696,7.567\right),\qquad\text{for }\gamma=1.
\end{equation}
$a_{\alpha}(\gamma)$ has a similar asymptotic behavior with $\gamma$
as the coefficient $c\left(\gamma\right)$ for the conductivity. For
$N=1$, $a_{1}=-3.756$ and $a_{2}=14.796$.

The solution of the kinetic equation has the form 
\begin{equation}
\chi_{\lambda,ij}(\mathbf{k})=I_{ij}\left(a_{1}+\beta\varepsilon_{\lambda,\mathbf{k}}a_{2}\right).\label{Chi3}
\end{equation}
The different components of the shear viscosity tensor are 
\begin{equation}
\eta_{ijk\ell}=\sum_{\lambda}\int_{\mathbf{k}}(v_{\lambda,\mathbf{k}})_{i}k_{j}\beta f_{\lambda}^{0}\left(1-f_{\lambda}^{0}\right)\chi_{\lambda,k\ell}(\mathbf{k}),\label{eta0}
\end{equation}
with $\eta_{xyxy}=\eta_{xyyx}=\eta_{yxxy}=\eta_{yxyx}\equiv\frac{3}{4}\eta_{0},$$\,\eta_{xzxz}=\gamma^{-2}\eta_{0},\,\eta_{zxzx}=\eta_{0},\,\eta_{xzzx}=\gamma^{-1}\eta_{0},$
and $\eta_{xzyz}=\eta_{xzzy}=\eta_{yzxz}=\eta_{yzzx}=0$ for the remaining
ones, where 
\begin{equation}
\eta_{0}(\gamma=1)\equiv N\kappa_{0}\gamma^{-1}\left(\frac{1}{v_{F}\beta}\right)^{3}\frac{1}{16\pi}\left[a_{1}\frac{\pi^{2}}{6}+a_{2}\frac{9}{2}\zeta(3)\right]\approx0.759N\frac{(k_{B}T)^{3}}{\alpha^{2}v_{F}^{3}},\label{eta1}
\end{equation}
for $N=2$. The numerical prefactor is $\approx1.469$ for $N=1$.

\end{widetext}

\section{Renormalization Group Analysis}

\begin{figure*}
\begin{centering}
\includegraphics[scale=0.4]{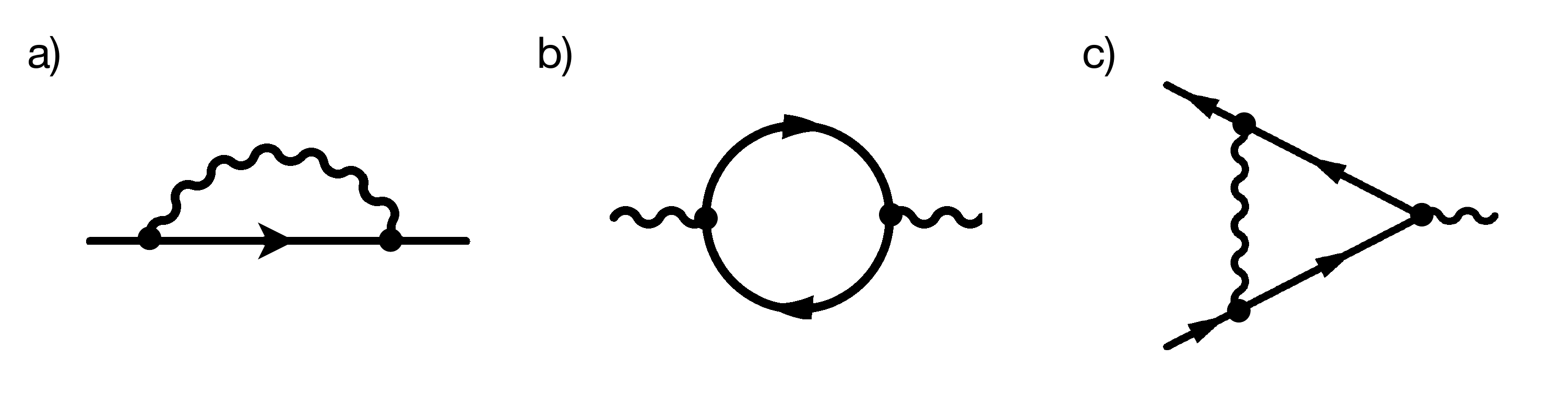}
\par\end{centering}
\caption{{\small{}\label{fig:appendix}(a) Self-energy, (b) polarization bubble,
and (c) vertex correction diagrams in one-loop perturbation theory.}}
\end{figure*}

We perform the renormalization group (RG) analysis using standard
perturbation theory. Since Coulomb interactions are marginal operators
in the RG sense, perturbation theory is well controlled in the regime
where the fine structure constant $\alpha=e^{2}/v_{F}\ll1$. In the
spirit of perturbation theory, in one loop one needs to extract the
leading logarithmic divergences of three diagrams: The Fock diagram
for the self-energy, the polarization bubble, and the vertex diagram,
as shown in Fig. \ref{fig:appendix}.

The Green's function for a NLSM is given by 
\begin{align}
\hat{G}^{-1}(i\nu,\mathbf{k}) & =i\nu-\frac{k_{r}^{2}-k_{F}^{2}}{2m}\sigma_{x}-v_{z}k_{z}\sigma_{y}\label{G}\\
 & \approx i\nu-v_{F}(\delta k_{r})\sigma_{x}-v_{z}k_{z}\sigma_{y},
\end{align}
with $v_{F}=k_{F}/m$ and $\delta k_{r}=k_{r}-k_{F}$. The pole of
the Green's function gives the energy dispersion 
\[
\pm\varepsilon(\mathbf{k})=\pm\sqrt{v_{F}^{2}(\delta k_{r})^{2}+v_{z}^{2}k_{z}^{2}},
\]
 whereas the Coulomb interaction is $V(q)=4\pi e^{2}/q^{2}.$

The Fock self-energy is given by the diagram 
\begin{equation}
\hat{\Sigma}(\mathbf{k})=-\frac{1}{\beta}\sum_{\nu}\int d^{3}k\,G(i\nu,\mathbf{k}+\mathbf{q})\frac{4\pi e^{2}}{q^{2}}.\label{Sigma2-1}
\end{equation}
At one loop level, the self-energy is frequency independent. In the
regime where the radius of the nodal line $k_{F}\gg\Lambda$, with
$\Lambda$ the momentum ultraviolet cut-off around the line, one can
ignore terms such as $q^{2}/k_{F}$, 
\begin{equation}
\frac{(\mathbf{k}+\mathbf{q})_{r}^{2}-k_{F}^{2}}{2m}\approx v_{F}(\delta k_{r}+\delta\hat{\mathbf{k}}_{r}\cdot\mathbf{q}_{r}).\label{approx}
\end{equation}

We integrate the bosonic momentum $q$ of the self-energy in the regime
$\delta k\ll q\ll k_{F}$, where the leading logarithmic divergence
of the diagram is expected.

Integrating in the frequency, it is convenient to calculate $\hat{\Sigma}(\mathbf{k})$
at $\mathbf{k}=(k_{F}+\delta k_{x},0,k_{z})$ and enforce rotational
symmetry around the nodal line, \begin{widetext}
\begin{align}
\hat{\Sigma}(k_{F}+\delta k_{x},0,k_{z}) & =\frac{1}{16\pi^{3}}\int_{-\Lambda}^{\Lambda}dq_{x}dq_{y}dq_{z}\frac{v(\delta k_{x}+q_{x})\sigma_{x}+v_{z}(k_{z}+q_{z})\sigma_{y}}{\varepsilon(\mathbf{k}+\mathbf{q})}\frac{4\pi e^{2}}{q^{2}}\nonumber \\
 & \stackrel{\delta k_{r}\ll q}{\longrightarrow}\frac{e^{2}}{4\pi^{2}}\int_{\delta k}^{\Lambda}d^{2}q_{\rho}\int_{-\Lambda}^{\Lambda}dq_{y}\frac{v_{z}^{2}vq_{z}^{2}\delta k_{x}\sigma_{x}+v_{z}v^{2}q_{x}^{2}k_{z}\sigma_{y}}{(v_{z}^{2}q_{z}^{2}+v^{2}q_{x}^{2})^{\frac{3}{2}}}\frac{1}{q_{\rho}^{2}+q_{y}^{2}}\nonumber \\
 & =\frac{e^{2}}{4\pi^{2}}\int_{0}^{2\pi}d\phi\,\frac{v_{z}^{2}v\cos^{2}\phi\,\delta k_{x}\sigma_{x}+v_{z}v^{2}\sin^{2}\phi k_{z}}{(v_{z}^{2}\cos^{2}\phi+v^{2}\sin^{2}\phi)^{\frac{3}{2}}}\ln\left(\frac{\Lambda}{\delta k}\right),\label{Sigma3}
\end{align}
with $q_{\rho}=\sqrt{q_{x}^{2}+q_{z}^{2}}$. The self-energy has the
form 
\begin{equation}
\hat{\Sigma}(k_{F}+\delta k_{x},0,k_{z})=\left[I_{1}(\gamma)v\delta k_{x}\sigma_{x}+I_{2}(\gamma)v_{z}k_{z}\sigma_{y}\right]\alpha\ln\left(\frac{\Lambda}{\delta k}\right)\label{Sigma2}
\end{equation}
\end{widetext}with $\gamma=v_{z}/v_{F}$, where 
\begin{align}
I_{1}(\gamma) & \equiv\frac{1}{4\pi}\int_{0}^{2\pi}d\phi\,\frac{\gamma^{2}\cos^{2}\phi}{[(\gamma^{2}-1)\cos^{2}\phi+1]^{\frac{3}{2}}}\label{I1}\\
I_{2}(\gamma) & \equiv\frac{1}{4\pi}\int_{0}^{2\pi}d\phi\,\frac{\sin^{2}\phi}{[(\gamma^{2}-1)\cos^{2}\phi+1]^{\frac{3}{2}}}\label{I2}
\end{align}
are elliptic integrals.

The perturbative velocity renormalization is 
\begin{align}
v & =v_{0}\left(1+\alpha_{0}I_{1}(\gamma)\ln\left(\frac{\Lambda}{\delta k}\right)\right),\label{v}\\
v_{z} & =v_{z0}\left(1+\alpha_{0}I_{2}(\gamma)\ln\left(\frac{\Lambda}{\delta k}\right)\right).\label{vz}
\end{align}

Next, we examine the vertex and the bubble diagrams. In standard perturbation
theory for Coulomb interactions, the vertex diagram does not contribute
to the charge renormalization due to a Ward identity, which relates
the vertex and the quasiparticle residue renormalizations. In one
loop, the self-energy is frequency independent, and hence the vertex
diagram is zero at this order. The polarization bubble renormalizes
the Coulomb interaction and could also renormalize the charge. However,
the static polarization bubble of a NLSM is perfectly regular and
does not contain logarithmic divergences \citep{Huh}, 
\begin{equation}
\Pi(0,q_{r},q_{z})\approx-\frac{N}{(2\pi)^{3}}\frac{k_{F}}{v_{F}q}\left(a_{1}q_{r}^{2}+a_{2}q_{z}^{2}\right),\label{Pi2}
\end{equation}
with $a_{1}$ and $a_{2}$ of order unity. Therefore, neither diagram
contributes to the renormalization of the charge, which does not run
in the perturbative regime.

We also point out that since the polarization $\Pi(0,q_{r},q_{z})$
is linear in $\mathbf{q}$ whereas $V(q)\propto1/q^{2}$, $\Pi$ changes
the form of the Coulomb propagator due to screening effects at small
$q$, 
\begin{equation}
\frac{4\pi e^{2}}{q^{2}-4\pi e^{2}\Pi(0,q_{r}q_{z})}.\label{screened}
\end{equation}
For $q\ll\frac{N\alpha}{2\pi^{2}}k_{F}=q_{c}$, Coulomb interactions
are screened (although still long range) and the analysis in the vicinity
of the fixed point will change. Our analysis indicates that further
away from that fixed point, for $q\gtrsim q_{c}$ and $N\alpha<1$,
where interactions are unscreened and standard perturbation theory
applies, only the velocities run. At low momentum, for $q\ll q_{c}$
and $N\alpha<1$ where interactions are partially screened, no logarithmic
divergences are present and the RG flow stops, whereas the charge
remains unrenormalized.

\subsection{Perturbative RG equations}

From Eqs. (\ref{v}) and (\ref{vz}), the corresponding RG equations
for the velocities are: 
\begin{align}
\frac{d\ln v}{d\ell} & =\alpha I_{1}(\gamma),\label{Rg1}\\
\frac{d\ln v_{z}}{d\ell} & =\alpha I_{2}(\gamma)
\end{align}

One can equivalently write two equivalent equations, 
\begin{align}
\frac{d\ln\gamma}{d\ell} & =\alpha\left[I_{2}(\gamma)-I_{1}(\gamma)\right]\approx\alpha\frac{1-\gamma}{8},\label{RG2}\\
\frac{d\ln\alpha}{d\ell} & =-\alpha I_{1}(\gamma)\approx-\frac{\alpha}{4}
\end{align}
In this regime, $\alpha$ runs towards an isotropic fixed point with
$\alpha=0$ and $\gamma=1$. The solution of the RG equations for
$\alpha$ and $\gamma$ is 
\begin{equation}
\alpha(\delta k)=\frac{\alpha_{0}}{1+\frac{\alpha_{0}}{4}\ln\left(\frac{\Lambda}{\delta k}\right)},\label{RG3}
\end{equation}
and 
\begin{equation}
\gamma^{-1}(\delta k)=1+\frac{\gamma_{0}^{-1}-1}{\left[1+\frac{\alpha_{0}}{2}\ln\left(\frac{\Lambda}{\delta k}\right)\right]^{\frac{1}{4}}},\label{gamma-1}
\end{equation}
while the velocity runs as 
\begin{equation}
v(\delta k)=v_{0}\left[1+\frac{\alpha}{4}\ln\left(\frac{\Lambda}{\delta k}\right)\right],\label{RGv2}
\end{equation}
as in graphene \citep{Kotov}.

\section{Derivation of the viscosity in hydrodynamic regime}

In a momentum conserved system, the continuity equation for momentum
is 
\begin{equation}
\frac{\partial\zeta_{j}\left(\mathbf{x},t\right)}{\partial t}+\partial_{i}T_{ij}\left(\mathbf{x},t\right)=0\label{eq:Cont}
\end{equation}
where $\zeta_{j}\left(x,t\right)$ is the momentum density in space
and time $x,t$. Indices $i$, $j$ refer to spatial components in
$d$ dimensions. The stress tensor operator $\tau_{ij}$ plays an
important role in the transport of viscous quantum fluids. $T_{ij}=-P_{i}\delta_{ij}+T_{ij}^{\prime}$
is composed of pressure $\mathbf{P}$ and of the viscous stress tensor
$T_{ij}^{\prime}$, which is the off diagonal part of the stress tensor
and can be defined as the expectation of the stress tensor due to
strain\cite{Landau}. \textcolor{black}{In non-equilibrium systems,
the deviation in the average stress tensor $\left\langle T_{\mu\nu}^{\prime}\right\rangle $
depends on the strain tensor and its time derivative in linear response,
\begin{equation}
\left\langle T_{ij}^{\prime}\right\rangle =\lambda_{ijk\ell}\xi_{k\ell}+\eta_{ijk\ell}\frac{\partial\xi_{k\ell}}{\partial t}\label{eq:dtau3}
\end{equation}
The component of the viscosity tensor $\eta_{ijk\ell}$ where the
component $i=j$ is called }\textit{\textcolor{black}{bulk viscosity}}\textcolor{black}{.
We are interested in the shear viscosity, where $i\neq j$, so we
use $T_{ij}$ and $T_{ij}^{\prime}$ interchangeably. Comparing classical
and quantum fluids, there is an analogous relation between the gradients
of the velocity field}\textbf{\textcolor{black}{{} $\mathbf{u}$}}\textcolor{black}{{}
and the time derivative of the metric tensor $\xi_{ij}$ \citep{Bradlyn}:
\begin{equation}
\frac{\partial u_{i}}{\partial x_{j}}=\frac{\partial\xi_{ij}}{\partial t}.\label{eq:strain}
\end{equation}
Thus, the shear viscosity can be obtained by the non-equilibrium stress
tensor, which is linearized with respect to space derivative of average
velocity $\vec{u}$.}

To find an effect of strain in the Hamiltonian in linear response,
we use the strain generator 
\begin{align}
\mathcal{J}_{ij} & =-\frac{1}{2}\sum_{n}\left\{ x_{i}^{n},p_{j}^{n}\right\} ,
\end{align}
where $n$ stands for particle indices. Following Bradlyn and Read's
approach at zero magnetic field \citep{Bradlyn}, the correction in
the Hamiltonian up to first order in $\xi_{\mu\nu}\left(t\right)$
can be shown to be 
\begin{equation}
\mathcal{H}_{1}=-\frac{\partial\xi_{ij}}{\partial t}\mathcal{J}_{ij}.\label{eq:Strain}
\end{equation}
In order to relate the total strain generator $\mathcal{J}_{ij}$
to the energy-stress tensor $\left\langle T_{ij}\right\rangle $,
we define the momentum density for a system of $n=1,2,\dots$ particles
in the absence of strain as 
\begin{equation}
\zeta_{i}\left(\mathbf{x},t\right)=\frac{1}{2}\sum_{n}\left\{ p_{i}^{(n)},\delta\left(x_{i}-x_{i}^{(n)}\right)\right\} ,\label{zeta}
\end{equation}
and then use the continuity equation (\ref{eq:Cont}) in momentum
representation, 
\begin{equation}
\partial_{t}\zeta_{i}\left(\mathbf{q},t\right)=-iq_{j}\tau_{ij}\left(\mathbf{q},t\right).\label{dtzeta1}
\end{equation}
Upon expanding the momentum density for small momentum $\mathbf{q}$,
we find $\zeta_{i}\left(\mathbf{q},t\right)$ as 
\begin{align}
\zeta_{i}\left(\mathbf{q},t\right) & =\int_{\mathbf{x}}e^{i\mathbf{q\cdot x}}\zeta_{i}\left(\mathbf{x},t\right)\nonumber \\
 & =\zeta_{i}\left(0,t\right)+iq_{j}\frac{1}{2}\sum_{n}\left\{ p_{i}^{(n)},x_{j}^{(n)}\right\} +\cdots
\end{align}
where $\zeta_{i}\left(0,t\right)$ is the direct momentum. Hence,
\begin{equation}
\partial_{t}\zeta_{i}\left(\mathbf{q},t\right)-\partial_{t}\zeta_{i}\left(0,t\right)=-\partial_{t}\left[iq_{j}\frac{1}{2}\left\{ x_{j},p_{i}\right\} \right]=\partial_{t}iq_{j}\mathcal{J}_{ij}.\label{Dtzeta2}
\end{equation}
If we set $\partial_{t}\zeta_{i}\left(0,t\right)=0$ due to global
momentum conservation and compare (\ref{dtzeta1}) and (\ref{Dtzeta2}),
the stress tensor is 
\begin{equation}
T_{ij}\left(\mathbf{q},t\right)=-\frac{\partial\mathcal{J}_{ij}}{\partial t}.
\end{equation}
When we write it in terms of quasiparticle operators, 
\begin{align}
T_{ij}\left(\mathbf{q},t\right) & =\sum_{\lambda,a}\int_{\mathbf{k}}\gamma_{\lambda,a}^{\dagger}\left(\mathbf{q}\right)\lambda\frac{\partial}{\partial t}\left(-\mathcal{J}_{ij}\right)\gamma_{\lambda,a}\left(\mathbf{q}\right)\nonumber \\
 & =\sum_{\lambda,a}\int_{\mathbf{k}}\gamma_{\lambda,a}^{\dagger}\left(\mathbf{q}\right)\frac{\lambda}{2}\frac{\partial}{\partial t}\left(x_{\mu}q_{\nu}+q_{\nu}x_{\mu}\right)\gamma_{\lambda,a}\left(\mathbf{q}\right),
\end{align}
and take the expectation value, then 
\begin{align}
\left\langle T_{\mu\nu}\right\rangle  & =\sum_{\lambda,a}\int_{\mathbf{k}}\lambda v_{\mu}q_{\nu}\left\langle \gamma_{\lambda,a}^{\dagger}\gamma_{\lambda,a}\right\rangle \label{eq:energystress}\\
 & =N\sum_{\lambda}\int_{\mathbf{k}}v_{\lambda,\mu}q_{\nu}f_{\lambda}\left(k,t\right),
\end{align}
with $N$ being the fermionic degeneracy.

\end{document}